%% file: main.tex
\documentclass[10pt,nonatbib]{sigplanconf}

\usepackage{microtype}
\usepackage{graphicx}
\usepackage{float}
\usepackage{booktabs}
\usepackage{multirow}
\usepackage{etoolbox}

\usepackage{algorithmic}
\usepackage{amssymb}
\usepackage{pifont}
\usepackage{listings}

\usepackage{color}
 
\newcommand{\cm}{\ding{51}}%
\usepackage[normalem]{ulem}
\usepackage[hyphens]{url}

\begin{document}

\CopyrightYear{2017} 
\setcopyright{acmcopyright}
\conferenceinfo{ASPLOS '17,}{April 08-12, 2017, Xi'an, China}
\isbn{978-1-4503-4465-4/17/04}\acmPrice{\$15.00}
\doi{http://dx.doi.org/10.1145/3037697.3037719}

\title{TriCheck: Memory Model Verification at the Trisection of Software, Hardware, and ISA}

\authorinfo{Caroline Trippel \enskip Yatin A. Manerkar \enskip Daniel Lustig* \enskip Michael Pellauer* \enskip Margaret Martonosi}
{Princeton University \and *NVIDIA}
{\{ctrippel,manerkar,mrm\}@princeton.edu \and \{dlustig,mpellauer\}@nvidia.com}

\maketitle

\begin{abstract}
\textit{Memory consistency models} (MCMs) which govern inter-module interactions in a shared memory system, are a significant, yet often under-appreciated, aspect of system design. MCMs are defined at the various layers of the hardware-software stack, requiring thoroughly verified specifications, compilers, and implementations at the interfaces between layers.
Current verification techniques evaluate segments of the system stack in isolation, such as proving compiler mappings from a high-level language (HLL) to an ISA or proving validity of a microarchitectural implementation of an ISA.

This paper makes a case for full-stack MCM verification and provides a toolflow, TriCheck, capable of verifying that the HLL, compiler, ISA, and implementation collectively uphold MCM requirements.
The work showcases TriCheck's ability to evaluate a proposed ISA MCM in order to ensure that each layer and each mapping is correct and complete.
Specifically, we apply TriCheck to the open source RISC-V ISA~\cite{RISCV}, seeking to verify accurate, efficient, and legal compilations from C11. We uncover under-specifications and potential inefficiencies in the current RISC-V ISA documentation and identify possible solutions for each. As an example, we find that a RISC-V-compliant microarchitecture allows 144 outcomes forbidden by C11 to be observed out of 1,701 litmus tests examined. Overall, this paper demonstrates the necessity of full-stack verification for detecting MCM-related bugs in the hardware-software stack.

\end{abstract}


\input{01-intro.tex}
\input{02-motivating.tex}
\input{03-method.tex}

\input{04-riscv.tex}

\input{05-casestudy.tex}

\input{06-results.tex}
\input{07-future.tex}
\input{08-related.tex}
\input{09-conclusion.tex}
\input{10-ack.tex}

\bibliographystyle{plain}
\bibliography{references}

\end{document}

%% file: 01-intro.tex
\section{Introduction}
\label{sec:intro}
Modern computer systems employ increasing amounts of heterogeneity and specialization to achieve performance scaling at manageable power and thermal levels.  
Reaping the full benefits of these heterogeneous and parallel systems often requires the cores to be able to communicate through shared memory.  This in turn necessitates \textit{memory consistency models} (MCMs)~\cite{adve:tutorial,gharachorloo:release,lamport:sc}, which specify the values that can be legally returned when loads access shared memory.
MCMs are defined at the various layers of the hardware-software stack and are central to hardware and software system design.
Properly designed MCMs enable programmers to synchronize and orchestrate the outcomes of concurrent code. Poorly designed MCMs can underspecify inter-core communication and lead to unreliability.


Several categories of problems can arise when translating a program from a high-level language (HLL) into correct, portable, and efficient assembly code. These include: (1) ill-specified or difficult-to-support HLL requirements regarding memory ordering; (2) incorrect compilation or mapping of instructions from the HLL to the target ISA; (3) inadequate ISA specification; and (4) incorrect microarchitectural implementation of the ISA.
If any of these issues are present in the hardware-software stack, code compiled for a given ISA may produce incorrect results.



\begin{figure}
  \begin{minipage}[b]{0.57\linewidth}
    \centering
    \label{fig:corsdwicpp}
    \includegraphics[width=.9\textwidth]{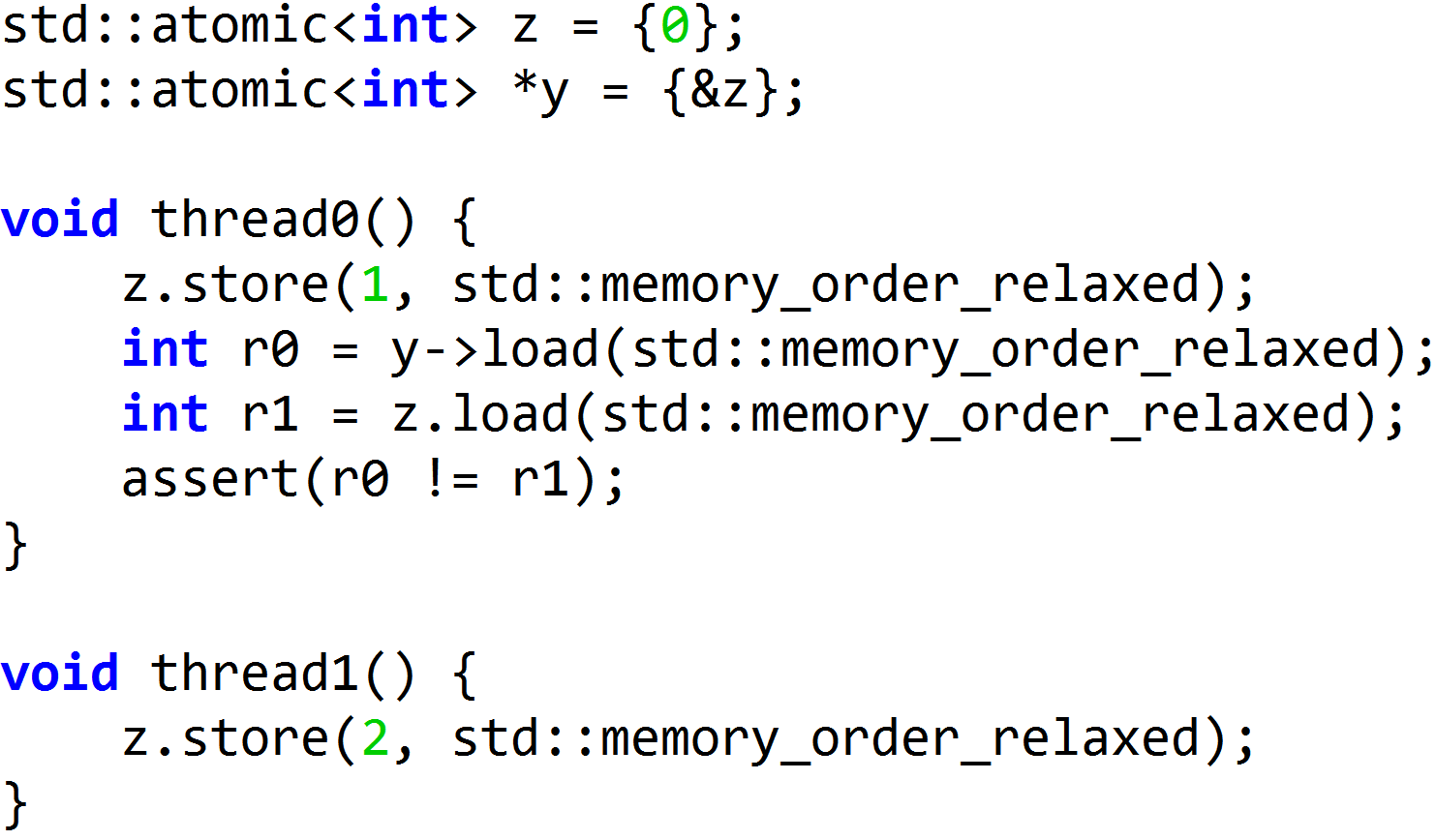}
  \end{minipage}%
  \begin{minipage}[b]{0.43\linewidth}
    \centering
    \label{fig:corsdwicppcycle}\includegraphics[width=.89\textwidth]{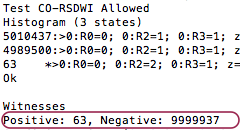}
  \end{minipage} 
  \caption{A C11 program that intermittently produces results disallowed by the C11 MCM when compiled by Clang++~v3.8 and run on modern ARM/Android hardware.}
  \label{fig:motivating}
\vspace{-5mm}
\end{figure}


Mis- and under-specification of MCMs in modern hardware is a real problem that leads to processors producing incorrect or counter-intuitive outcomes~\cite{ARMHazard}.
Consider the C11 program in Figure~\ref{fig:corsdwicpp}. When compiled by Clang++ v3.8, the resulting program \textit{intermittently} produces a result that is illegal according to the C11 specification~\cite{cppconcurrency} when run on \textit{some} ARM hardware platforms. This behavior was first reported by Alglave et al.\@~\cite{alglave:herd}. We have observed the phenomenon on a Galaxy Nexus (ARM Cortex-A9) and a Nexus 6 (Qualcomm Snapdragon 805).
In this particular example, the illegal outcome occurs because hardware does not preserve program order (i.e., the original thread ordering) for reads of the same address. This behavior was formally acknowledged by ARM as a bug in 2011~\cite{ARMHazard}, and is henceforth referred to in this paper as the ARM load$\rightarrow$load hazard.

The ARM load$\rightarrow$load hazard arose because the ARM ISA specification was ambiguous regarding the required ordering of same address loads, leading some implementations to relax the ordering. A precise ISA MCM specification is central to facilitating accurate translation from HLLs to assembly programs and implementing hardware that can correctly execute these programs.
If the ISA MCM is unclear, or if its definition is fundamentally at odds with the requirements of the HLL\footnote{Throughout this paper we will focus on the C11/C++11 HLL MCM, as it is widely applicable and rigorously defined~\cite{problemconcurrency}.} it intends to support, there is no longer a verifiable interface for compilers to target and for hardware to implement.

When errors do arise, their causes may be debated. Regardless of where blame is assigned, designers may propose a solution that affects other layers of the hardware-software stack out of convenience or necessity. In this case, due to the existence of buggy microarchitectures in the wild and the relative maturity of the ARM ISA, ARM elected to solve the problem in the compiler by stipulating that additional fences be added~\cite{ARMHazard}.
This paper advocates for MCMs as first-class citizens in the design of hardware-software ecosystems. It makes the following contributions:
\begin{itemize}

    \item 
    We present TriCheck, a framework for full-stack MCM verification. We demonstrate how TriCheck can aid system designers in verifying that HLL, compiler, ISA, and implementation align well on MCM requirements.  In particular, TriCheck supports iteratively designing an ISA MCM that provides an accurate and minimally-constrained target for compiled HLL programs.
    Our verification methodology systematically compares the {\em language-level} executions of HLL programs with their corresponding {\em ISA-level} executions on {\em microarchitectural implementations} of the ISA in question.
    When a microarchitectural execution differs from its corresponding language-level execution in a way that is illegal, TriCheck provides information that aids designers in determining if the cause is an incorrect compiler mapping, ISA specification, hardware implementation, or even HLL specification in some cases (see Section~\ref{sec:future}).
    
    
    \item We apply TriCheck to the open-source RISC-V ISA~\cite{RISCV} to validate TriCheck's applicability to modern ISA design. In particular, we assess the accuracy, precision,  and completeness of the specified RISC-V MCM in serving as a compiler target for C11 programs. Our work finds gaps in the RISC-V MCM specification. In particular, for a suite of 1,701 litmus tests, we present a microarchitecture that is compliant with the RISC-V specification yet incorrectly allows 144 outcomes forbidden by C11 to be observed.
    
    \item Based on the results of our evaluation, we propose improvements to the RISC-V ISA and MCM specification, in order to address the model's current shortcomings. A formalization of our proposal in Alloy~\cite{jackson:software,wickerson:comparing} is ongoing work.
    
    \item We showcase the benefits of full-stack MCM verification to areas other than ISA design by discussing the use of TriCheck to find two counterexamples \cite{manerkar:compilermappings} to the supposedly proven-correct trailing-sync compiler mappings from C11 to the Power and ARMv7 architectures \cite{c++topower}. 
\end{itemize}

%% file: 02-motivating.tex
\section{Background: Features of Memory Models}
\label{sec:motivating}
MCMs specify the rules and guarantees governing the ordering and visibility of accesses to shared memory.
Frequently regarded as the most intuitive MCM, Sequential Consistency (SC)~\cite{lamport:sc} requires that the result of a program execution is the same as if all cores execute their own instructions in program order (PO), and a total global order exists on all instructions from all cores such that each load returns the value written by the most recent store to the same address.
Unfortunately, common microarchitectural optimizations violate SC, resulting in low performance for naive SC implementations. In hardware, there have been many attempts at mitigating SC's performance cost, commonly leveraging techniques such as aggressive post-retirement speculation and rolling back execution in the case of a coherence violation~\cite{invisifence, bulksc, speculativesc, sc_ilp_rc, Ranganathan:speculativeretirement, Wenisch:storewaitfree}. Additionally, techniques have been proposed that aim to enforce SC only for conflicting accesses~\cite{atomicsc,efficientsc,endtoendsc}.
Nevertheless, most manufacturers have elected to build hardware with MCMs that relax SC. 
Various issues can arise when the effects of relaxing memory orderings are not carefully considered at ISA design time. Here we provide some examples of MCM features that are relevant to our case study and results in Sections~\ref{sec:casestudy} and~\ref{sec:results}.

\subsection{Coherence and Same-Address Ordering}\label{sec:coherence}
\textit{Coherence} ensures that (1) all stores are eventually made visible to all cores and (2) there exists a single total order that all threads agree on for all stores to the same address ~\cite{gharachorloo:thesis, gharachorloo:release}. \textit{Consistency} can be thought of as a superset of coherence in that it is additionally concerned with orderings of accesses to different addresses.
Accesses from the same thread to the same address generally must maintain program order, 
but there are exceptions: some old Power models and SPARC RMO relax same-address load$\rightarrow$load ordering~\cite{SPARCRMO,POWER4}.

\begin{figure}[t]
  \centering
  \includegraphics[width=0.45\textwidth]{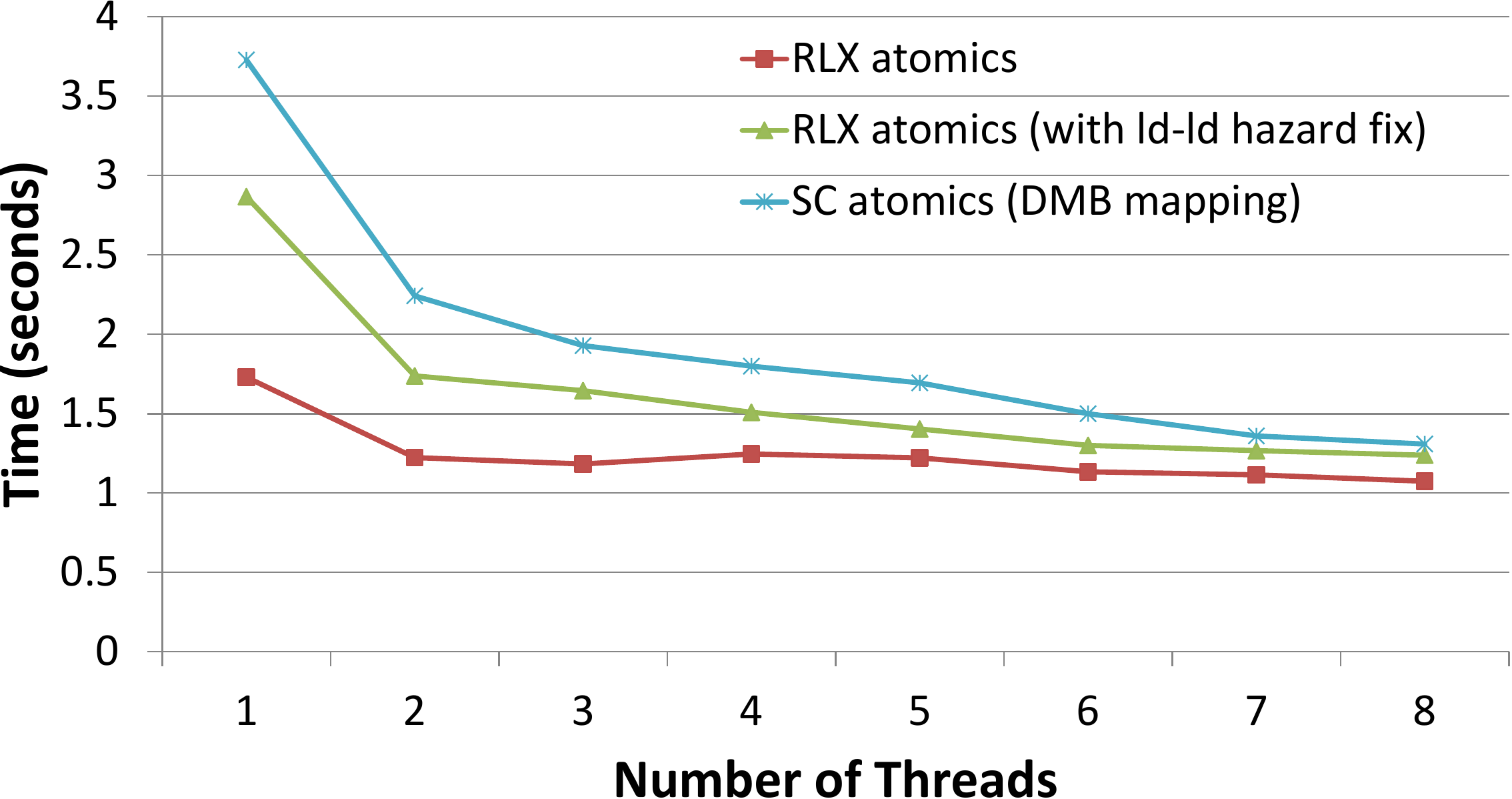}
  \caption{Runtimes of three variants of the Parallel Sieve of Eratosthenes~\cite{Boehm05} on an 8-core Samsung Galaxy S7 for up to 8 threads.}
  \label{fig:dmboverhead}
\end{figure}

Notably, imprecision in the coherence specification led to the ARM load$\rightarrow$load hazard discussed in Section~\ref{sec:intro}.
ARM acknowledged that due to the vast number of load instructions in programs, binary patching in the linker is infeasible; they instead suggest that compilers be rewritten to issue a \texttt{dmb} fence instruction immediately following atomic~\footnote{C11 uses ``atomic'' to mean ``memory accesses used for synchronization'', not just for read-modify-write.} loads.
As one demonstration of the cost of imprecise ISA MCM specifications, we estimate the overhead of this workaround using the parallel Sieve of Eratosthenes algorithm~\cite{Boehm05}.  This application gives the same results regardless of whether there is any synchronization between threads.  Thus, its reading and marking of entries can be implemented with either relaxed atomics or sequentially consistent atomics without compromising correctness.



We implemented three variants of the parallel sieve algorithm and recorded their runtimes for a problem size of $10^8$ on a Samsung Galaxy S7 with an Exynos 8890 8-core processor. Figure~\ref{fig:dmboverhead} shows the run times for thread counts between 1 and 8. The first of the three variants uses relaxed atomics, which map to ordinary loads and stores on ARM. The second uses relaxed atomics with a \texttt{dmb} fence added after relaxed loads, in accordance with ARM's recommended fix for the hazard. The third uses sequentially consistent (SC) atomics (implemented by surrounding the relevant stores with \texttt{dmb} fences in addition to placing \texttt{dmb} fences after the relevant loads---the standard ARM recipe).


The relaxed variant with the fix is always slower than the uncorrected relaxed atomic variant; this is due to the extra \texttt{dmb} fence after relaxed loads. The overhead of the fix is 15.3\% additional execution time at 8 threads. Furthermore, the performance of the fixed variant degrades to the level of fully sequentially consistent atomics at 8 threads. This experiment indicates that the overhead of fixing the load$\rightarrow$load hazard can be quite significant. 
We revisit the issue of same address load-load ordering in the context of the RISC-V MCM in Section~\ref{sec:ld_ld_reorder_riscv}.


\begin{figure}[t]
\begin{center}
\centering
\small
\setlength\tabcolsep{4pt} 
    \begin{tabular}{| c  c  c |}
    \hline
    \multicolumn{3}{|c|}{\textbf{Intial conditions}: x=0, y=0}\\ \hline
    T0                 & T1                 & T2                 \\ \hline
    a: st(x,1,rlx) & b: r0 = ld(x, rlx) & d: r1 = ld(y, acq) \\
                       & c: st(y,1,rel) & e: r2 = ld(x, rlx) \\ \hline
    \multicolumn{3}{|c|}{\textbf{Forbidden C11 Outcome}: r0=1, r1=1, r2=0}     \\
    \hline
    \end{tabular}
\end{center}
\caption{C11 variant of the Write-to-Read Causality (WRC) litmus test. T0, T1, and T2 are three threads. The st and ld of y perform release-acquire synchronization. }
\label{fig:wrc}
\end{figure}

\begin{figure}[t]
\begin{center}
\centering
\small
\setlength\tabcolsep{4pt} 
    \begin{tabular}{| c  c  c  c |}
    \hline
    \multicolumn{4}{|c|}{\textbf{Initial conditions}: x=0, y=0}\\ \hline
    T0            & T1            & T2            & T3    \\ \hline
    a: st(x,1,sc) & b: st(y,1,sc) & c: r0 = ld(x,sc) & e: r2 = ld(y,sc) \\
                  &               & d: r1 = ld(y,sc) & f: r3 = ld(x,sc) \\ \hline
    \multicolumn{4}{|c|}{\textbf{Forbidden C11 Outcome}: r0=1, r1=0, r2=1, r3=0}     \\
    \hline
    \end{tabular}
\end{center}
\caption{C11 variant of the Independent Reads of Independent Writes (IRIW) litmus. All accesses are SC atomics.}
\label{fig:iriw}

\end{figure}

\subsection{Dependencies}
  \label{sec:dependencies}
    A dependency relates a load with a load or store that is later in program order (PO).
    An \textit{address} dependency results when the address accessed by a load or store depends syntactically\footnote{ARM and Power in particular respect syntactic dependencies, which define dependencies according to the syntax of the instructions.  This is broader than semantic dependencies, which only include true dependencies, i.e., those which could not in theory be optimized away.} on the value returned by a PO-prior load.
    A \textit{data} dependency exists between a load and a PO-later store when the store's value depends syntactically on the loaded value.
    A \textit{control} dependency occurs when the control flow decision of whether to execute a load or store depends syntactically on the value returned by a PO-prior load.
    Intuitively, it may seem impossible not to enforce dependencies, as a dependee seemingly cannot execute until it has all of its inputs available.
    However, in the presence of microarchitectural speculation, the dependee can in fact behave as if it were reordered with the instruction it depends on~\cite{Martin:valueprediction}, unless such behavior is explicitly prevented by the ISA specification.

\subsection{Store Atomicity, Cumulativity, and C11 Atomics}
    \label{sec:atomicity}
    \subsubsection{Store Atomicity}
    As defined by Collier, a store is \textit{multiple-copy atomic} (MCA) if all cores in the system, including the performing core, conceptually see the updated value at the same instant~\cite{collier}.
    As a performance optimization, some architectures allow a core to read its own writes prior to their being made visible to other cores; we refer to this as \textit{read-own-write-early-multiple-copy atomic} (rMCA)~\cite{adve:tutorial}.
    However, rMCA writes must be made visible at the same time to all cores other than the performing core.
    Weaker models, like ARM and Power, feature \textit{non-multiple-copy atomic} (nMCA) stores that may become visible to some remote cores before they become visible to others.

    Figure~\ref{fig:wrc} demonstrates the often counter-intuitive effects of nMCA stores.
    The specified non-SC outcome corresponds to a causality chain where T0 sets a flag by writing 1 to \texttt{x}, and T1 reads the updated value of \texttt{x}, subsequently setting its own flag by writing 1 to \texttt{y}.
    T2 then sees the update of \texttt{y}, reading 1; however, it has still not observed the update of \texttt{x} and reads its value as 0. 
    If this C11 program is compiled down to regular loads and stores on a nMCA system, the forbidden outcome will (perhaps surprisingly) be observable.

    C11 supports cross-thread synchronization via acquire and release operations. These operations were initially proposed as part of release consistency (RC)~\cite{gharachorloo:release}. An acquire ensures that it is made visible before accesses after the acquire in program order. Likewise, a release ensures that accesses before it in program order are made visible before the release. The store and load of y in Figure~\ref{fig:wrc} form \textit{a release-acquire pair} that synchronizes the values between T1 and T2. C11 additionally requires release-acquire synchronization to be \textit{transitive}~\cite{Batty:mathematizingc++,cppconcurrency}. This means that T2 must observe the store to x when it acquires y, because T1 observed the store to x before its release of y. As a result, the outcome in Figure~\ref{fig:wrc} is forbidden.
    
    
    \subsubsection{Cumulativity}

\begin{table}
\begin{center}
\centering
\small
\setlength\tabcolsep{4pt} 
\begin{tabular}{| c | c | c | c | c | c |}

 \hline
 
 \textbf{C/C++ Instruction} & \textbf{Power}        \\ \hline
 \texttt{ld rlx}            & ld                    \\ \hline
 \texttt{ld acq}            & ld; ctrlisync         \\ \hline
 \texttt{ld sc}             & hwsync; ld; ctrlisync \\ \hline
 \texttt{st rlx}            & st                    \\ \hline
 \texttt{st rel}            & lwsync;st             \\ \hline
 \texttt{st sc}             & hwsync; st            \\
 \hline
 
\end{tabular}
\end{center}
\caption{Leading-sync compiler mapping from Power to C11~\cite{mckenney}.}
\label{table:powermappings}
\end{table}
    An nMCA architecture must include \textit{cumulative} fences in order to support C11-style cross-thread synchronization.
    Fences order specified accesses in the fence's \textit{predecessor set} (i.e., accesses before the fence) with specified accesses in the fence's \textit{successor set} (i.e., accesses after the fence)\footnote{Predecessor and successor sets are called \textit{group A} and \textit{group B} in descriptions of fences in the Power and ARM memory models~\cite{sarkar2011}.}.    
    Cumulative fences additionally include accesses performed by threads other than the fencing thread in the predecessor and successor sets. Recursively, memory operations (from any thread) that have performed prior to an access in the predecessor set are also members of the predecessor set. Also recursively, memory operations (from any thread) that perform after a load that returns the value of a store in the successor set are also in the successor set.
    \subsubsection{C11 Compiler Mappings}
    \label{sec:c11compilermappings}
    The C11 memory model has various forms of synchronization with different strength/performance trade-offs, and so ISAs often provide a corresponding set of synchronization primitives with similar trade-offs.
    As C11 release-acquire synchronization is transitive, it requires cumulative ordering at the hardware-level between a release-acquire pair. Compiler mappings for well-known nMCA architectures such as Power and ARMv7 enforce all cumulative ordering requirements of the C11 release-acquire pair on the release side, leaving the acquire side implementation non-cumulative. In this case, the cumulative fence on the release side would require that reads and writes in the predecessor set be ordered with writes in the successor set, and that reads in the predecessor set be ordered with reads in the successor set (i.e., \textit{cumulative lightweight fence}). Power mappings implement release operations using a similar cumulative lightweight fence (\texttt{lwsync}). With all cumulative orderings being enforced on the release side, acquire operations only need to locally order reads before the fence with reads and writes after the fence. Power and ARMv7 can implement acquire operations using non-cumulative fences (e.g., \texttt{ctrlisync} and \texttt{ctrlisb}, respectively\footnote{\texttt{ctrlisync} and \texttt{ctrlisb} represent the \texttt{cmp;~bc;~isync} and \texttt{teq;~beq;~isb} instruction sequences, respectively.}). We adopt the Power approach, which we provide in Table~\ref{table:powermappings}, in our proposed mappings for the RISC-V ISA in Section~\ref{sec:riscv}.

    C11 also supports sequentially consistent (SC) atomics. An SC load is an acquire operation and an SC store is a release operation, and there must also be a total order on all SC atomic operations that is respected by all cores. As such, the program in Figure~\ref{fig:iriw} must forbid the listed outcome, as there is no total order of SC operations that would allow it. At the architecture level, cumulative lightweight fences as described for release-acquire synchronization are not sufficient to implement the required ordering for this program. Even if a cumulative lightweight fence was placed between each pair of loads on T2 and T3, neither T2 nor T3 reads from writes after the fences, so the writes observed before the fences need not be propagated to other cores. Instead, fences used to implement C11 SC atomics must be cumulative fences that order reads and writes before the fence with reads and writes after the fence (i.e., \textit{cumulative heavyweight fence}).
    Power and ARMv7 use cumulative heavyweight fences (\texttt{sync} and \texttt{dmb}, respectively) to implement C11 SC atomics.

%% file: 03-method.tex
\section{Full-Stack MCM Verification}
\label{sec:method}

TriCheck is the first tool capable of full stack MCM verification bridging the HLL, compiler, ISA, and microarchitecture levels. MCMs are defined at the various layers of the hardware-software stack, and errors at any layer or in translating between layers can produce incorrect results. No other tool can run this top-to-bottom analysis, and TriCheck does so efficiently enough to find real bugs.



\subsection[Background: Check and Herd]{Background: \textit{Check} and \textit{Herd}}
The TriCheck approach builds on the \textit{Check}~\cite{checkweb,pipecheck,coatcheck, armor, ccicheck} family of tools. 
A hardware designer can use a domain-specific language (DSL) called \textit{$\mu$Spec} to describe a microarchitecture by defining a set of ordering axioms. This specification along with a collection of user-provided litmus tests and corresponding required outcomes for each test serve as inputs to Check tools. Check tools evaluate the correctness of the processor model by comparing the required litmus test outcomes with outcomes that are \textit{observable} on the model. Furthermore, Check enables designers to model speculative execution as well as the subtle interplay between coherence and consistency~\cite{ccicheck} and virtual memory and consistency~\cite{coatcheck}.

TriCheck uses these tools along with \textit{Herd}~\cite{alglave:herd}, a MCM simulator that takes as input a user-defined MCM (in a concise format) and a litmus test and outputs all executions of that tests that are \textit{permitted} by the model. In contrast with the Check tools, Herd defines more abstract axiomatic models that do not depend on microarchitectural details.  
Recent work has added support for language-level MCMs, and in particular, a model has been constructed for C11~\cite{batty:overhauling}, which we use in our case study in Section~\ref{sec:casestudy}.

\subsection{The TriCheck Methodology}
\label{sec:trichecksteps}
\begin{figure}[t]
    \centering

    
    {\includegraphics[width=0.9\linewidth]{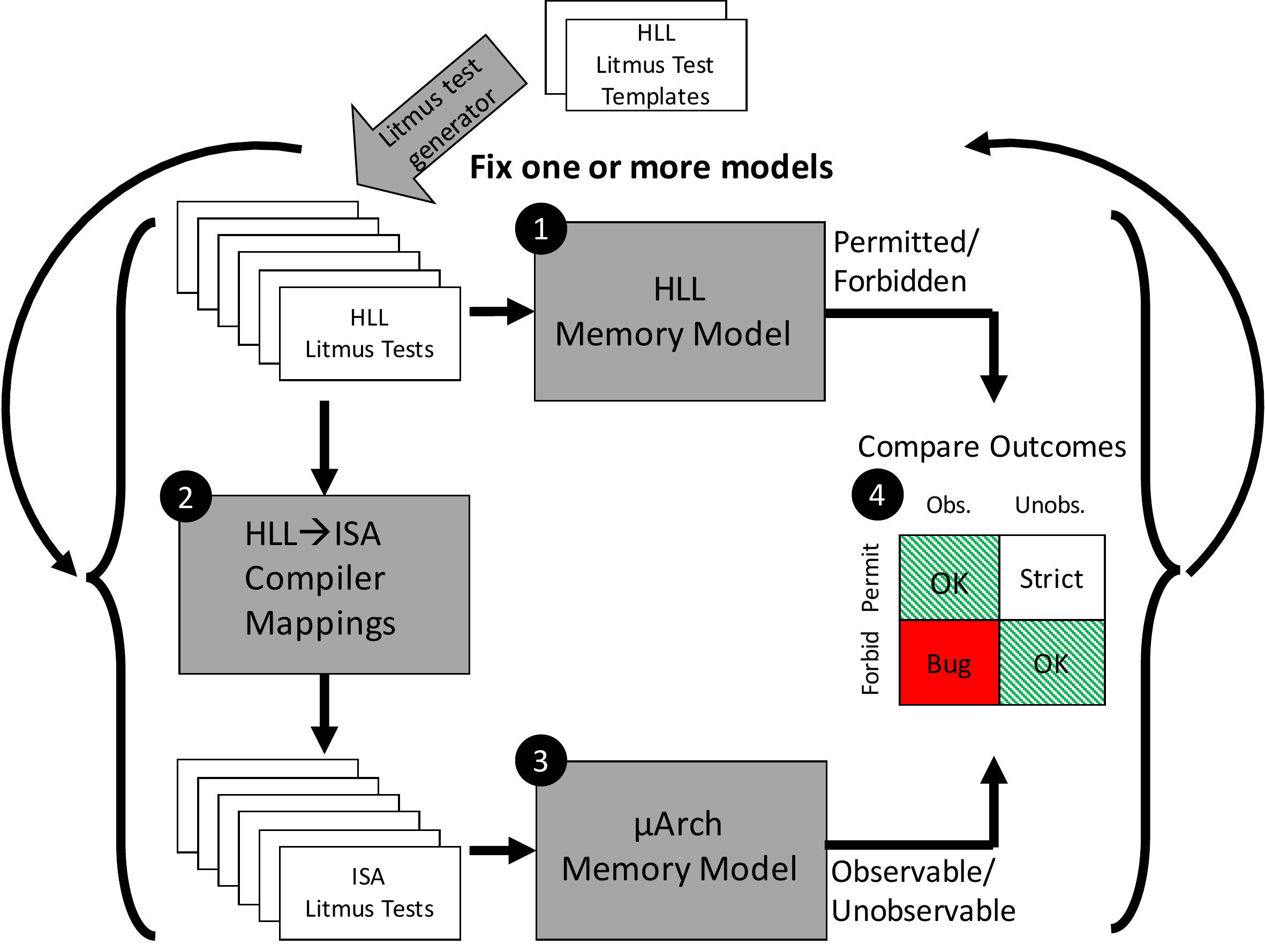}}
    %
    \caption{TriCheck toolflow for full-stack MCM verification. Bugs may require modified ISA or HLL MCMs, different sets of enforced orderings from either the compiler or the microarchitecture, or more or fewer ISA instructions with specified ordering semantics. Numbers correspond to TriCheck steps enumerated in Section~\ref{sec:trichecksteps}.}
    \label{fig:method}
    \vspace{-4.5pt}
\end{figure}

The ISA MCM serves as a contract between hardware and software. It defines ordering semantics of valid hardware implementations and provides ordering-enforcement mechanisms for compilers to leverage.
We identify four primary MCM-dependent system components: a HLL MCM, compiler mappings from the HLL to an ISA, an ISA MCM, and a microarchitectural implementation of the ISA.
In Figure~\ref{fig:method}, we illustrate the TriCheck framework and include as inputs a HLL MCM, HLL$\rightarrow$ISA compiler mappings, an implementation model, and a suite of HLL litmus tests. The ISA places constraints on both the compiler and the microarchitecture and is present in TriCheck via these two inputs.
Given these inputs, TriCheck evaluates whether or not they can successfully work together to preserve MCM ordering semantics guaranteed to the programmer when HLL programs are compiled and run on target microarchitectures.



We envision architects using TriCheck early in the ISA or microarchitecture design process.
While architects are selecting hardware optimizations for improved performance or simplifications for ease of verification, TriCheck can be used to simultaneously study the effects of these choices on the ISA-visible MCM and the ability of their designs to accurately and efficiently support HLL programs.

However, TriCheck is not limited to new or evolving ISA designs. Similar to the ARM load$\rightarrow$load hazard in Section~\ref{sec:motivating}, there are cases when other elements (e.g., the compiler) are modified in response to ISA or microarchitecture MCM bugs out of convenience or necessity. When a solution is proposed for a MCM bug---such as fence insertion in ARM's case---TriCheck can be used to verify that adding the fence did indeed prohibit the forbidden outcome across relevant litmus tests.
Our case study in Section~\ref{sec:casestudy} showcases TriCheck's applicability to ISA design by focusing on the time in the design process when the ISA MCM can be modified. Section~\ref{sec:future} describes ongoing work where we are using TriCheck to evaluate compiler mappings.

\begin{figure}
    \centering
    {\includegraphics[width=0.9\linewidth]{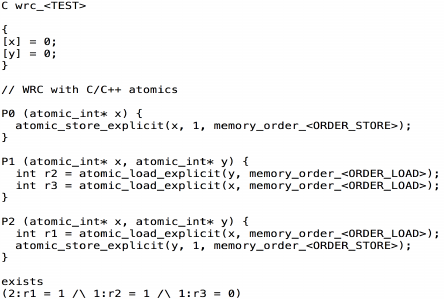}}
    %
    \caption{Example of a C11 Herd litmus test template for the WRC litmus test.}
    \label{fig:wrctemplate}
\end{figure}

TriCheck is a litmus-test-based verification framework. To get the best coverage, TriCheck should consider a variety of interesting tests. We provide a \textit{litmus test generator} capable of producing a suite of interesting tests from litmus test templates containing placeholders that correspond to different types of memory or synchronization operations, and a set of HLL MCM primitives to insert into these placeholders. An example template is shown in Figure~\ref{fig:wrctemplate}. With the C11 MCM for example, litmus test templates would contain placeholders for memory reads, writes, and/or fences. Our litmus test generator would then produce all permutations of each test template, featuring all combinations of applicable C11 \texttt{memory\_order} primitives. This allows us to verify ISA MCM functionality for all possible \texttt{memory\_order} interactions and synchronization scenarios for a given litmus test. Other methods of litmus test generation are possible, and in particular, a new tool has been designed to produce the optimal set of litmus tests for a given memory model~\cite{lustig:automated}.

Figure~\ref{fig:method} depicts the TriCheck toolflow. The gray shaded boxes represent TriCheck's non-litmus-test \textsc{inputs}: a HLL MCM specification, compiler mappings from HLL MCM primitives to ISA assembly instructions, and a $\mu$Spec model corresponding to an implementation of the ISA MCM.

In describing the TriCheck toolflow, we will discuss how the inputs are combined, evaluated and can be refined in order to prohibit all illegal-according-to-HLL executions and to permit as many legal-according-to-HLL executions as possible. (This results in correct but minimally constrained HLL programs.)
Given its inputs, the TriCheck toolflow proceeds as follows:
\begin{enumerate}
    \item \label{item:axiomatic} \textsc{HLL Axiomatic Evaluation}: The suite of HLL litmus tests is run on a HLL Herd model (e.g., the C11 Herd model) to determine the outcomes that the HLL MCM \textit{permits} or \textit{forbids} for each at the program level.
    \item \label{item:compilation} \textsc{HLL$\rightarrow$ISA Compilation}: Using the HLL$\rightarrow$ISA compiler mappings, TriCheck translates HLL litmus tests to their assembly language equivalents.
    \item \label{item:operational} \textsc{ISA $\mu$Spec Evaluation}: The suite of assembly litmus tests is run on the Check model of the ISA to determine the outcomes that are \textit{observable} or \textit{unobservable} on the microarchitecture.
    \item \label{item:equivalence} \textsc{HLL-Microarchitecture Equivalence Check}: The results of Step~\ref{item:axiomatic} and Step~\ref{item:operational} are compared for each test to determine if the microarchitecturally realizable outcomes are \textit{stronger than}, \textit{weaker than}, or \textit{equivalent to} the outcomes required by the HLL model. A \textit{stronger than} (resp. \textit{weaker than}) outcome corresponds to a HLL program that is \textit{permitted} (resp. \textit{forbidden}) by the HLL MCM, yet \textit{unobservable} (resp. \textit{observable}) on the microarchitectural implementation of the ISA. If Step~\ref{item:equivalence} concludes that the microarchitecturally realizable outcomes are more restrictive than what the HLL requires, the designer may wish to relax the ISA or microarchitecture for performance reasons. On the other hand, some correction is \textit{mandatory} when outcomes forbidden by the HLL are observable on the microarchitecture.
\end{enumerate}

After running TriCheck on a combination of inputs, a subsequent \textsc{Refinement} step is possible. This step corresponds to refining any combination of the HLL MCM, compiler mappings, and microarchitectural implementation. This refinement step is the process of modifying an input in response to a microarchitectural execution that differs from the HLL-specified outcome for a given set of program executions. The purpose of refinement is to have a better match between HLL requirements and execution outcomes, whether this is to eliminate bugs or avoid overly-constraining the implementation.



%% file: 04-riscv.tex
\section{Case Study: The RISC-V ISA MCM}
\label{sec:riscv}
The RISC-V ISA is a free, open RISC ISA.
A widely-utilized, free and open ISA offers some key advantages, such as well-maintained compiler and software tool-chains and even open-source hardware designs.
However, a clearly defined ISA MCM is crucial in achieving this vision.
To demonstrate the applicability of our framework to modern ISA design, we have conducted a case study that applies our toolflow from Section~\ref{sec:method} to the latest version of the RISC-V ISA specification~\cite{RISCV}.

In this experiment, we study both the \textit{Baseline} (labeled ``Base'') and the \textit{Baseline + Atomics Extension} (labeled ``Base+A'') RISC-V ISAs, evaluating each on how efficiently and accurately they are able to (or \textit{not} able to) serve as compiler targets for C11 programs.
For example, the program in Figure~\ref{fig:wrc} can show the forbidden outcome when compiled to the RISC-V ISA (via the ``Intuitive" compiler mappings detailed in Tables~\ref{table:compilermappings} and \ref{table:compilermappingsAMO}) if the microarchitectural implementation leverages nMCA stores (which the RISC-V specification allows). \textit{For the Base ISA, we show that there is no way to provide a correct mapping.}
For the Base+Atomics ISA, we show that likely-unintended inefficiencies can result from modifying the mapping to force the correct C11-required outcome.

\subsection {RISC-V Baseline MCM}
\begin{table}
\begin{center}
\centering
\footnotesize
\setlength\tabcolsep{3pt} 
\begin{tabular}{| c | c | c |}

  \cline{2-3}
  \multicolumn{1}{c}{}&\multicolumn{2}{| c |}{\textbf{ C11 $\rightarrow$ RISC-V Base Compiler Mappings}}       \\\hline 
  \textbf{C11 Atomics} & \textbf{Intuitive}       & \textbf{Refined} \\\hline
  \texttt{ld rlx}        & ~~~ ld ~~~                  & ld                   \\ \hline
  \texttt{ld acq}        & ~~~ ld;f[r,m] ~~~          & ld;f[r,m]            \\ \hline
  \texttt{ld sc}         & ~~~ f[m,m];ld;f[m,m] ~~~    & hwf;ld;f[r,m]  \\ \hline
  \texttt{st rlx}        & ~~~ st ~~~                   & st                   \\ \hline
  \texttt{st rel}        & ~~~ f[m,w]; st ~~~          & lwf;st   \\ \hline
  \texttt{st sc}         & ~~~ f[m,m]; st ~~~          & hwf;st          \\ \hline
 
\end{tabular}
\end{center}
\caption{\textit{Intuitive} and \textit{Refined} compiler mappings from C11 to RISC-V Base ISA. \textit{f[a,b]} is a fence that orders type \textit{a} accesses before and \textit{b} accesses after the fence. \textit{lwf} and \textit{hwf} are the cumulative lightweight and heavyweight fences from Section~\ref{sec:c11compilermappings}. \textit{r}, \textit{w}, and \textit{m} are reads, writes, and memory operations, respectively. \textit{Refined} mappings are presented in Section~\ref{sec:casestudy}.}
\label{table:compilermappings}

\end{table}

\label{sec:riscv_baseline}
\subsubsection{Relaxed memory model} The Base RISC-V MCM (Section 2.7 of the RISC-V ISA specification~\cite{RISCV}) allows multiple threads of execution within a single user address space that may communicate and synchronize via the shared memory system. Each thread must observe \textit{its own} memory operations as if they executed sequentially in program order. However the manual specifies that RISC-V has a ``relaxed memory model'' that requires explicit \texttt{FENCE} instructions to guarantee any specific ordering between memory operations as viewed by other RISC-V threads.

\subsubsection{\texttt{ \bf FENCE} Instructions for Memory Accesses} RISC-V allows any combination of memory read and write instructions may be ordered with any combination of the same. The manual states that, ``Informally, no other RISC-V thread or external device can observe any operation in the successor set following a \texttt{FENCE} before any operation in the predecessor set preceding the \texttt{FENCE}.'' 
We interpret predecessor and successor sets here to be the accesses of the specified type(s) that come before and after the \texttt{FENCE} in program order, respectively.

\subsubsection{Dependencies} Of particular note is the fact RISC-V does not require memory ordering to be enforced for dependent instructions, even though this can result in counter-intuitive outcomes in multiprocessor systems~\cite{Martin:valueprediction}.
Many commercial architectures such as x86, ARM, and Power respect address, data, and some control dependencies between instructions, and such dependencies can also be used as lightweight synchronization to enforce orderings locally~\cite{sarkar2011}.
More importantly, if dependency orderings are not preserved by default, they must be explicitly enforced through ISA instructions when necessary. For example, the Linux kernel includes a \texttt{read\_barrier\_depends()} barrier that is used to conditionally enforce data dependencies on systems that do not respect them, such as Alpha~\cite{linuxbarrier}.
We note that the current Linux port of RISC-V does not map \texttt{read\_barrier\_depends()} to any fence, and so may be incorrect for some microarchitectural implementations~\cite{linuxriscvbarrier}. \underline{Our recommendation} is to require the preservation of dependency orderings in the ISA memory model.
Other issues with the RISC-V memory model are discussed in Section~\ref{sec:casestudy}.

\subsection{RISC-V Baseline + Atomics Extension}

\begin{table}
\begin{center}
\centering
\footnotesize
\setlength\tabcolsep{3pt} 
\begin{tabular}{| c | c | c |}

  \cline{2-3}
  \multicolumn{1}{c}{}&\multicolumn{2}{| c |}{\textbf{ C11 $\rightarrow$ RISC-V Base+A Compiler Mappings}}       \\\hline
  \textbf{C11 Atomics} & \textbf{Intuitive}     & \textbf{Refined}     \\\hline
  \texttt{ld rlx}            & ~~~~~~~~  ld ~~~~~~~~    & ld                   \\ \hline
  \texttt{ld acq}            & ~~~~~~~ AMO.aq ~~~~~~~~ & AMO.aq               \\ \hline
  \texttt{ld sc}             & ~~~~~~~~ AMO.aq.rl ~~~~~~~~  & AMO.aq.sc            \\ \hline
  \texttt{st rlx}            & ~~~~~~~~ st ~~~~~~~~     & st                   \\ \hline
  \texttt{st rel}            & ~~~~~~~~ AMO.rl ~~~~~~~~ & AMO.rl               \\ \hline
  \texttt{st sc}             & ~~~~~~~~ AMO.aq.rl ~~~~~~~~  & AMO.rl.sc            \\ \hline
 
\end{tabular}
\end{center}
\caption{\textit{Intuitive} and \textit{Refined} compiler mappings from C11 to RISC-V Base+A ISA. \textit{AMO.a} is an AMO operation with the \textit{a} bit set, etc. \textit{Refined} mappings are presented in Section~\ref{sec:casestudy}.}
\label{table:compilermappingsAMO}

\end{table}
\label{sec:riscv_baseline_atomics}
\subsubsection{RMWs with memory orders} The \textit{Standard Extension for Atomic Instructions} (Chapter 6 of the RISC-V ISA specification \cite{RISCV}) contains atomic fetch-and-op instructions (i.e., AMOs) and Load-Reserve/Store-Conditional (LR/SC) instructions. Both of these read-modify-write mechanisms may be annotated with various memory ordering semantics---\textit{unordered}, \textit{acquire}, \textit{release}, and \textit{sequentially consistent}. The manual states that these ordering mechanisms are meant to ``implement the C11 and C++11 memory models efficiently.'' They are defined as follows:
\begin{itemize}
    \item \textit{Unordered}: ``No additional ordering constraints are imposed on the atomic memory operation.''
    \item \textit{Acquire (Acq)}: ``No following memory operations on this RISC-V thread can be observed to take place before the Acq memory operation.'' The manual also states that \texttt{FENCE~R,~RW} suffices to implement acquire orderings.
    \item \textit{Release (Rel)}: ``The Rel operation cannot be observed to take place before any earlier memory operations on this RISC-V thread.'' The manual also states that \texttt{FENCE~RW,~W} suffices to implement release orderings.
    \item \textit{Sequentially Consistent (SC)}: ``The SC operation is sequentially consistent and cannot be observed to happen before any earlier memory operations or after any later memory operations in the same RISC-V thread, and can only be observed by any other thread in the same global order of all sequentially consistent atomic memory operations to the same address domain.''
\end{itemize}

\subsubsection{Store atomicity} The manual states that nMCA implementations are allowed, but that for SC operations, the specification requires ``full sequential consistency for the atomic operation which implies global store atomicity in addition to both acquire and release semantics.''

\subsection{Microarchitectural Implementations}
\label{sec:uarch_explanation}

To support our evaluation of the RISC-V MCMs with TriCheck, we implemented in $\mu$Spec a set of microarchitectures (summarized in Table~\ref{table:relaxations}) that relax various aspects of program order and store atomicity while remaining RISC-V-compliant.
We constructed these models by extending a model of the RISC-V Rocket Chip~\cite{rocketchip}, a 6-stage in-order pipeline that supports the Base RISC-V ISA and some optional extensions, including the Atomics extension.
These models were augmented with the appropriate RISC-V instructions depending on whether they were implementing the Base or Base+A ISA.
The ordering variations we study are:
\begin{enumerate}
    \item \textbf{WR}: W$\rightarrow$R reordering is achieved by buffering stores in a FIFO queue prior to eventually pushing them out to the rest of the memory hierarchy. Value forwarding is disallowed, but younger loads may complete when their effective address does not match the address of any earlier store still in the store buffer. 
    \item \textbf{rWR}: Builds on WR by allowing value forwarding from stores in the store buffer to later loads of the same address.
    \item \textbf{rWM}: Extends rWR by allowing writes (to different addresses) to retire from the store buffer out of order. Coherence requires a total global order on stores to the same address.
    \item \textbf{rMM}: Extends rWM by allowing reads to commit out of order with earlier reads or writes. We maintain that read$\rightarrow$write ordering must be maintained for same address reads and writes, but we allow reordering for all read$\rightarrow$read pairs in this baseline version, including same address read$\rightarrow$read pairs.
    \item \textbf{nWR}: Extends rWR by allowing cores to share store buffers. This is analogous to having a shared write-through cache~\cite{cppconcurrency}, and allows nMCA stores.
    \item \textbf{nMM}: Extends rMM by allowing shared store buffers in the same vein as nWR.
    \item \label{a9_description} \textbf{A9like}: To demonstrate that the visibility of nMCA behavior does not depend on having a design that contains a shared buffer or shared write-through cache, we model another microarchitecture with ISA-visible relaxations that match those of nMM. This time we leverage the ability of Check to model subtleties of the cache-coherence/consistency interface. To implement nMCA stores, this model features: i) write-back caches that allow multiple requests for write permission (for different addresses) to be in progress at the same time and ii) a non-stalling directory coherence protocol that allows the storing core to forward the store's value to another core before it has received all invalidations for the access. In this scenario, coherence is preserved, but nMCA stores arise. This design captures reordering features similar to those allowed by the ARM Cortex-A9~\cite{armcortexa9}.
\end{enumerate}

\begin{figure}[t]
\begin{center}
\centering
\small
\setlength\tabcolsep{2pt} 
\begin{tabular}{| l | c | c | c | c | c | c |}
 \cline{2-7}
 \multicolumn{1}{c|}{} & \multicolumn{3}{c|}{\textbf{Relaxed PO}} & \multicolumn{3}{c|}{\textbf{Store Atomicity}} \\ \hline
 \multicolumn{1}{|l|}{\textbf{$\mu$Spec Model}} & W$\rightarrow$R & W$\rightarrow$W & R$\rightarrow$M & MCA & rMCA & nMCA \\ \hline
 WR     & \cm   &       &       & \cm   &       &       \\ \hline
 rWR    & \cm   &		&       &       & \cm   &       \\ \hline
 rWM	& \cm   & \cm   &       &       & \cm   &       \\ \hline
 rMM	& \cm	& \cm	& \cm   &       & \cm   &       \\ \hline
 nWR    & \cm   &       &		&       &       & \cm   \\ \hline
 nMM    & \cm	& \cm   & \cm	&       &    	& \cm   \\ \hline
 A9like & \cm	& \cm   & \cm	&       &    	& \cm*  \\ 
 \hline
 
\end{tabular}
\end{center}
\caption{$\mu$Spec models that relax various aspects of program order and store atomicity while remaining RISC-V-compliant~\cite{adve:tutorial}. Section~\ref{sec:uarch_explanation} (Point~\ref{a9_description}) discusses the difference between A9like and nMM. }
\label{table:relaxations}
\end{figure}

%% file: 05-casestudy.tex
\section{Applying our Methodology to RISC-V}
\label{sec:casestudy}

As a case study of our approach, we use TriCheck to analyze the RISC-V ISA's memory models.
We divide our case study into two halves, one for the Base ISA model and one for the Base+A ISA model.
For each of these specifications, we begin with the MCM as specified in Sections \ref{sec:riscv_baseline} and \ref{sec:riscv_baseline_atomics} respectively. Our initial compiler mappings are the ``Intuitive'' mappings from Table~\ref{table:compilermappings}. These mappings are derived from information in the RISC-V manual~\cite{RISCV}.
For the microarchitecture component of our analysis, we utilize the microarchitectures detailed in Section~\ref{sec:uarch_explanation} (augmented with instructions unique to the Base or Base+A ISA as appropriate), starting with the strongest---WR.

We apply the iterative design and refinement methodology of Figure~\ref{fig:method} to these inputs. 
When bugs are encountered, we propose a solution and re-run TriCheck to confirm the fix is successful. Additionally, we incrementally explore weaker and weaker microarchitectures from Table~\ref{table:relaxations}, pushing the bounds of what the RISC-V MCMs allow.
Our analysis shows that parts of the current RISC-V MCMs are \textit{too weak} and others are \textit{too strong} to implement C11 atomics correctly and efficiently. We recommend a set of possible model refinements to fix their problems, and use our framework to ensure that these changes have the desired effect.

\subsection{Base RISC-V Model Analysis \& Refinement}
\label{sec:base_analysis}

The Base ISA only provides memory fence instructions to establish synchronization between threads. As such, C11 atomics must be implemented in the Base ISA using a combination of fences and ordinary loads and stores. 

\subsubsection{Lack of Cumulative Lightweight Fences}
\label{sec:base_lack_cum}

\begin{figure}[t]
\begin{center}
\centering
\small
\setlength\tabcolsep{3pt} 
    \begin{tabular}{| c c  c  c |}
    \hline
    \multicolumn{4}{|c|}{\textbf{Initial conditions}: x=0, y=0}\\ \hline
    ~~~~T0            &~~~~& T1            & T2            \\ \hline
    ~~~~a: sw x1, (x5) &~~~~& b: lw x2, (x5) & e: lw x3, (x6) \\
                   &~~~~& c: fence rw, w & f: fence r, rw \\
                       &~~~~& d: sw x2, (x6) & g: lw x4, (x5) \\ \hline
    \multicolumn{4}{|c|}{\textbf{Forbidden C11 Outcome}: x1=1, x2=1, x3=1, x4=0}     \\
    \hline
    \end{tabular}
\end{center}
\caption{Figure~\ref{fig:wrc} WRC variant compiled to RISC-V Base using Table~\ref{table:compilermappings} ``Intuitive" compiler mappings. Registers x5 and x6 hold the addresses of x and y respectively on all cores.}
\label{fig:wrc_analysis}
\end{figure}

\begin{figure}[t]
\begin{center}
\centering
\small
\setlength\tabcolsep{1.5pt} 
    \begin{tabular}{| c  c  c  c |}
    \hline
    \multicolumn{4}{|c|}{\textbf{Initial conditions}: x=0, y=0}\\ \hline
    T0            & T1            & T2            & T3    \\ \hline
    a: fence rw, rw &  c: fence rw, rw & e: fence rw, rw & k: fence rw, rw \\
    b: sw x1, (x7) & d: sw x2, (x8) & f: lw x3, (x7) & l: lw x5, (x8) \\
                &                   & g: fence rw, rw & m: fence rw, rw \\
                &                   & h: fence rw, rw & n: fence rw, rw \\
                &                    & i: lw x4, (x8) & o: lw x6, (x7) \\
                &                   & j: fence rw, rw & p: fence rw, rw \\ \hline
    \multicolumn{4}{|c|}{\textbf{Forbidden C11 Outcome}: x1=1, x2=1, x3=1, x4=0, x5=1, x6=0}     \\
    \hline

    \end{tabular}
\end{center}
\caption{Figure~\ref{fig:iriw} IRIW variant compiled to RISC-V Base using Table~\ref{table:compilermappings} ``Intuitive" compiler mappings. Registers x7 and x8 hold the addresses of x and y respectively on all cores.}
\label{fig:iriw_analysis}
\end{figure}

As covered in Section~\ref{sec:atomicity}, C11 release-acquire synchronization is required to be transitive, ordering both accesses before a release in program order \textit{and} accesses that were observed by the releasing core prior to the release. As such, in the WRC variant of Figure~\ref{fig:wrc}, it is forbidden for T2 to return 0 for its load of x if it observes the release to y using its acquire. This ordering is not implicitly enforced for regular loads and stores in nMCA memory systems, which RISC-V allows.

When we ran the Base MCM through TriCheck using the ``Intuitive'' compiler mappings from Table~\ref{table:compilermappings}, the test in Figure~\ref{fig:wrc} compiled down to that in Figure~\ref{fig:wrc_analysis}. Analysis of this program with Check indicated that the forbidden outcome was observable on the microarchitecture. Upon investigation of the results, we deduced that the bug was due to the absence of cumulative fences in the Base ISA.

The Base RISC-V ISA does not contain \textit{any} cumulative fences that are capable of enforcing this ordering. Thus, this problem \textit{cannot} be fixed simply by changing the compiler mapping. \underline{Our recommended solution} to the problem is to modify the ISA such that the fences used to implement releases are cumulative, specifically \textit{cumulative lightweight fences} as defined in Section~\ref{sec:motivating}.

We modified the microarchitectural implementation of the fences used for releases to be cumulative lightweight fences, and reran TriCheck with the new microarchitecture. This time, the tests such as WRC that require cumulative lightweight orderings disallowed the forbidden outcomes.

\subsubsection{Lack of Cumulative Heavyweight Fences}
\label{sec:base_lack_strong_cum}

As discussed in Section~\ref{sec:atomicity}, the enforcement of a total order is necessary for C11 SC atomics. This requirement is exhibited by the variant of the IRIW litmus test shown in Figure~\ref{fig:iriw}, whose non-SC outcome is forbidden by C11.

Using the ``Intuitive" compiler mappings from Table~\ref{table:compilermappings}, the test compiles down to Figure~\ref{fig:iriw_analysis} for the Base ISA. Check reported that the forbidden outcome was allowed by our microarchitectural implementation for this test. Examination of the graph generated by Check showed that this was also due to the lack of cumulativity in fences. However, unlike the WRC case above, cumulative lightweight fences between the pairs of loads on T2 and T3 are \textit{insufficient} to enforce the ordering required, and we verified this using TriCheck. Instead, as discussed in Section~\ref{sec:atomicity}, cumulative heavyweight fences are required to prohibit the forbidden outcome in this case - a feature which the Base ISA does not provide.

As in Section~\ref{sec:base_lack_cum}, \underline{our recommended solution} to this problem is to modify the ISA to include \textit{cumulative heavyweight fences}. We modified the microarchitectural implementation to support cumulative heavyweight fences, changed the compiler mappings to use these fences when mapping sequentially consistent atomics, and reran the modified setup through TriCheck. We observed that the forbidden outcome of Figure~\ref{fig:iriw} was disallowed with the new instructions and mapping.

\subsubsection{Reordering Loads to the Same Address}
\label{sec:ld_ld_reorder_riscv}
After making the above changes and rerunning TriCheck on the modified setup, we observed that variants of the CoRR and CO-RSDWI litmus tests were still producing forbidden outcomes. These bugs were occurring because the microarchitectural implementation was not ordering loads to the same address (an ordering that RISC-V does not require). As discussed in Section~\ref{sec:motivating}, C11 atomics require that two loads of the same address maintain program order. The ``Intuitive'' compiler mapping for relaxed atomics from Table~\ref{table:compilermappings} implements relaxed atomics with regular loads and stores, which implies that the microarchitecture should enforce this ordering requirement; however, the microarchitecture was not doing so because the Base ISA does not require this. As a result, the forbidden outcome is visible on the microarchitecture.

This issue can be fixed in one of two ways: either the compiler mapping for C11 relaxed loads can be changed to add a fence after each, or the ISA memory model can be modified to require loads to the same address to be ordered by hardware. As a relatively new ISA, RISC-V can use either option. However, adding fences after each relaxed load can result in significant performance degradation for programs that liberally use relaxed atomics, as seen in Section~\ref{sec:coherence}. As such, a more efficient solution is for the ISA memory model to require program order to be preserved between two loads to the same address. We modified the microarchitecture to provide this ordering and used TriCheck to verify that the forbidden outcome no longer occurred.

\subsection{RISC-V Base+Atomics Model Analysis \& Refinement}
\label{sec:base_atomics_analysis}
Virtually all of the instructions unique to Base+A  are read-modify-write (RMW) instructions.
The deficiencies in the Base model mentioned above apply to the Base+A MCM as well. However, analysis with TriCheck shows that the new instructions in Base+A cannot implement C11 atomic operations correctly and efficiently, as we detail below.

The RISC-V manual~\cite{RISCV} states that an atomic load operation may be implemented as an AMOADD to the zero register and an atomic store operation can be implemented as an AMOSWAP operation that writes the old value to the zero register (in other words, by discarding the store and load portions of certain RMWs).

\subsubsection{Lack of Cumulative Releases}
\label{sec:lack_cum_releases}

\begin{figure}[t]
\begin{center}
\centering
\footnotesize
\setlength\tabcolsep{2pt} 
    \begin{tabular}{| c  c  c |}
    \hline
    \multicolumn{3}{|c|}{\textbf{Initial conditions}: x=0, y=0}\\ \hline
    T0            & T1            & T2            \\ \hline
    a: sw x1, (x5) & b: lw x2, (x5) & d: amoadd.w.aq x0, x3, (x6) \\
                       & c: amoswap.w.rl x2, x0, (x6) & e: lw x4, (x5) \\ \hline
    \multicolumn{3}{|c|}{\textbf{Forbidden C11 Outcome}: x1=1, x2=1, x3=1, x4=0}     \\
    \hline
    \end{tabular}
\end{center}
\caption{RISC-V Base+A version of Figure~\ref{fig:wrc} WRC variant using Table~\ref{table:compilermappingsAMO} ``Intuitive" compiler mappings. Registers x5 and x6 hold the addresses of x and y respectively on all cores.}
\label{fig:wrc_base_atomics}
\end{figure}

As discussed in Section~\ref{sec:atomicity}, C11 releases are required to be transitive by the C11 memory model, which necessitates cumulative fences. However, release instructions in the current RISC-V specification are \textit{not} required to be cumulative, and only order the accesses before them in program order. As a result, using the ``Intuitive'' compiler mapping for atomics in Table~\ref{table:compilermappingsAMO}, the test in Figure~\ref{fig:wrc} compiles down to that in Figure~\ref{fig:wrc_base_atomics}. Check analysis of this test indicates that the forbidden outcome is visible on the microarchitecture, signifying a bug.

Note that even if the compiler mapping were changed to use \textit{AMO.aq.rl} operations (the strongest synchronization instructions the ISA provides) for releases, the problem would persist. Even though AMO.aq.rl operations are store atomic and have both acquire and release semantics, they are not cumulative and will not enforce the required ordering (we verified this with TriCheck). 

\underline{Our recommended solution} to this issue is to make release operations in the RISC-V ISA cumulative, requiring that accesses before a release in program order \textit{and} writes observed by the releasing core before the release be made visible before the release is made visible. Using TriCheck, we verified that making these changes to the microarchitecture's implementation of releases resulted in the forbidden outcome of Figure~\ref{fig:wrc_analysis} being correctly disallowed.

Furthermore, if the AMO.rl.sc instruction is MCA and cumulative, then it is sufficient to implement an SC store (we verified this using TriCheck). This is because the cumulative release semantics ensure that all previous accesses (including previously observed writes) are made visible before the release, and the store atomicity of the release ensures that the release is made visible to all cores at the same time.


\subsubsection{Absence of Roach-Motel Movement for SC Atomics}
\label{sec:roach_motel}

\begin{figure}[t]
\begin{center}
\centering
\small
\setlength\tabcolsep{1pt} 
    \begin{tabular}{| c  c |}
    \hline
    \multicolumn{2}{|c|}{\textbf{Initial conditions}: x=0, y=0}\\ \hline
    ~~~~~~~~T0            & T1            \\ \hline
    ~~~~~~~~a: st(x,1,sc) & c: r0 = ld(y,sc) \\
    ~~~~~~~~b: st(y,1,rlx)     & d: r1 = ld(x,sc) \\ \hline
    \multicolumn{2}{|c|}{\textbf{Allowed C11 Outcome}: r0=1, r1=0}     \\
    \hline
    \end{tabular}
\end{center}
\caption{A variant of the C11 Message Passing (MP) litmus test where the store to \texttt{y} is a relaxed operation and can bypass the store to \texttt{x} through roach motel movement.}
\label{fig:mp_roach}
\end{figure}

\begin{figure}[t]
\begin{center}
\centering
\small
\setlength\tabcolsep{2pt} 
    \begin{tabular}{| c  c |}
    \hline
    \multicolumn{2}{|c|}{\textbf{Initial conditions}: x=0, y=0}\\ \hline
    T0            & T1            \\ \hline
    a: amoswap.w.aq.rl x1, x0, (x4) & c: amoadd.w.aq.rl x0, x2, (x5) \\
    b: sw x1, (x5) & d: amoadd.w.aq.rl x0, x3, (x4) \\ \hline
    \multicolumn{2}{|c|}{\textbf{Forbidden RISC-V Outcome}: x1=1, x2=1, x3=0}     \\
    \hline
    \end{tabular}
\end{center}
\caption{RISC-V Base+A version of Figure~\ref{fig:mp_roach} MP variant using Table~\ref{table:compilermappings} ``Intuitive" compiler mappings. Registers x4 and x5 hold the addresses of x and y respectively on both cores.}
\label{fig:mp_roach_atomics}
\end{figure}

In the C11 memory model, SC loads and stores only need to enforce acquire and release orderings respectively, in addition to appearing in a total order observed by all cores. SC loads do not need to implement release semantics and SC stores do not need to implement acquire semantics~\cite{cpp14}. As a result, ordinary loads and     stores (as well as relaxed atomics) that follow an SC store or precede an SC load in program order can be reordered before the SC store or after the SC load respectively. This is known as  \textit{roach-motel movement} and intuitively corresponds to making a critical section larger, which will not break code that uses atomic operations and locks in well-structured ways~\cite{cppconcurrency}. Roach motel movement allows acquires and releases to function as one-way barriers, allowing more reordering of memory operations and  theoretically improved performance.

The RISC-V ISA requires both the \texttt{aq} and \texttt{rl} bits to be set on an AMO operation in order to ensure the store atomicity required to correctly implement SC   operations. There is no way to have MCA operations with only acquire or release semantics, which would map closely to the requirements of C11 SC loads and     stores. As a result, the implementations of SC loads and stores in the RISC-V Base+A ISA is too strict, unnecessarily enforcing more orderings than the C11 model requires. For example, in the version of the \texttt{mp} litmus test shown in Figure~\ref{fig:mp_roach}, the C11 memory model allows the relaxed store to y to be reordered before    the SC store to x by roach motel movement. Thus, it is possible for T1 to observe the store to \texttt{y} before it observes the store to \texttt{x}.

However, when using the ``Intuitive'' RISC-V mapping from Table~\ref{table:compilermappingsAMO}, the program in Figure~\ref{fig:mp_roach} compiles down to that in Figure~\ref{fig:mp_roach_atomics}. When we ran this program through Check's implementation verification, it deduced that the allowed outcome was in fact forbidden by the microarchitecture. Specifically, the acquire semantics of the AMO.aq.rl used to implement the SC store to x \textit{prevents} the store to y from being reordered with it through roach motel movement.

One way to fix this excessive ordering enforcement is to decouple the store atomicity setting of an AMO from its acquire and release semantics, allowing AMOs to be store   atomic when only having acquire or release semantics. We denote such store atomic AMOs as AMO.\{aq$|$rl\}.sc. Using TriCheck, we verified that if SC loads are mapped to   AMO.aq.sc and SC stores are mapped to AMO.rl.sc, the outcome in Figure~\ref{fig:mp_roach} is allowed, and no forbidden outcomes are allowed as a result of this relaxation.

\subsubsection{Lazy Implementation of Cumulativity}
\label{sec:lazy_impl}

In the C11 memory model, acquire and SC loads (resp., acquire and SC fences) can only synchronize with release and SC stores (resp., release and SC fences). In other words, if a release is observed by a relaxed atomic access,   it is \textit{not} necessary for the thread performing the relaxed atomic access to observe all accesses before the release as well. It is only when the release is observed by an   acquire or an SC load that the accesses before the release must be observed by the loading core. As such, in the version of the \texttt{mp} litmus test shown in Figure~\ref{fig:mp_lazy}, it is valid for T1 to observe the store to \texttt{y} but then still return the old value of 0 for \texttt{x}. This is true even though the execution of the two loads on T1 is locally ordered through means of an address dependency (assuming dependencies are respected---see Section~\ref{sec:riscv_baseline}). Enforcing that releases only synchronize with acquire operations allows for ``lazy" implementations of cumulativity, which can delay processing coherence invalidations until an acquire operation is reached. Such implementations can help reduce false sharing and consume less bandwidth~\cite{elver14,hower14,keleher92}, and should not be outlawed by an ISA memory model specification  if possible.

The C11 constraints on the observation of a release are slightly different to the constraints on the observation of a release in RISC-V. In RISC-V, a release is considered to synchronize with respect to a given core when it is observed by \textit{any} load on that core, and not necessarily by an acquire. Using the ``Intuitive'' compiler mappings from Table~\ref{table:compilermappingsAMO}, the test in Figure~\ref{fig:mp_lazy} compiles down to the code in Figure~\ref{fig:mp_lazy_atomics}. The microarchitectural verification step in our framework confirmed that the allowed outcome was unnecessarily forbidden by the microarchitecture.

In order to allow this outcome and enable lazy, high-performance implementations of the RISC-V Base+A ISA, \underline {our recommend solution} is to modify the ISA to dictate that a release need only synchronize with respect to a core when it is observed by an acquire operation from that core. Upon making this modification to the microarchitectural implementation used in our analysis, we verified that the outcome from Figure~\ref{fig:mp_lazy} was now allowed by the microarchitecture.

\begin{figure}[t]
\begin{center}
\centering
\small
\setlength\tabcolsep{2pt} 
    \begin{tabular}{| c  c |}
    \hline
    \multicolumn{2}{|c|}{\textbf{Initial conditions}: x=0, y=0}\\ \hline
        ~~~~~~~~T0            & T1            \\ \hline
        ~~~~~~~~a: st(x,1,rel) & c: r0 = ld(y,rlx) \\
        ~~~~~~~~b: st(y,x,rel)     & d: r1 = ld(r0, acq) \\ \hline
    \multicolumn{2}{|c|}{\textbf{Allowed C11 Outcome}: r0=x, r1=0}     \\
    \hline
    \end{tabular}
\end{center}
\caption{A variant of the C11 Message Passing (MP) litmus test where the load of \texttt{y} is a relaxed operation and need not synchronize with the store to \texttt{y} on T0. Note the address dependency between the two instructions on T1.}
\label{fig:mp_lazy}
\end{figure}

\begin{figure}[t]
\begin{center}
\centering
\small
\setlength\tabcolsep{2pt} 
    \begin{tabular}{| c  c |}
    \hline
    \multicolumn{2}{|c|}{\textbf{Initial conditions}: x=0, y=0}\\ \hline
    T0            & T1            \\ \hline
    a: amoswap.w.rl x1, x0, (x4) & c: lw x2, (x5) \\
    b: amoswap.w.rl x4, x0, (x5) & d: amoadd.w.aq x0, x3, (x2) \\ \hline
    \multicolumn{2}{|c|}{\textbf{Forbidden RISC-V Outcome}: x1=1, x2=x, x3=0}     \\
    \hline
    \end{tabular}
\end{center}
\caption{RISC-V Base+A version of Figure~\ref{fig:mp_lazy} MP variant using Table~\ref{table:compilermappingsAMO} ``Intuitive" compiler mappings. Registers x4 and x5 hold the addresses of x and y respectively on both cores.}
\label{fig:mp_lazy_atomics}
\end{figure}

\subsection{Refined RISC-V Compiler Mappings}
\label{sec:refinemappings}
At the conclusion of our RISC-V case study, we arrived at the ``Refined'' compiler mappings from C11 to RISC-V Base and Base+A outlined in Tables~\ref{table:compilermappings} and \ref{table:compilermappingsAMO}. We note that this case study served to evaluate only Base and Base+A mappings in isolation (i.e., we did not evaluate the mixing of fences and AMO synchronization operations for supporting the C11 MCM). Similarly, ARMv7 mappings cannot inter-operate with ARMv8 mappings~\cite{sewell:mappings}. Furthermore, one could imagine different fence primitives for implementing C11 acquire and release operations that feature more symmetry in terms of ordering semantics (i.e., splitting cumulative ordering responsibilities between acquire and release operations). Our choice of fences here is meant to be as compatible as possible with the current RISC-V specification and instruction format.

%% file: 06-results.tex
\section{RISC-V MCM Shortcomings Quantified}
\label{sec:results}
\label{sec:data}
\begin{figure*}
    \centering
    {\label{fig:results81A}\includegraphics[width=.85\linewidth]{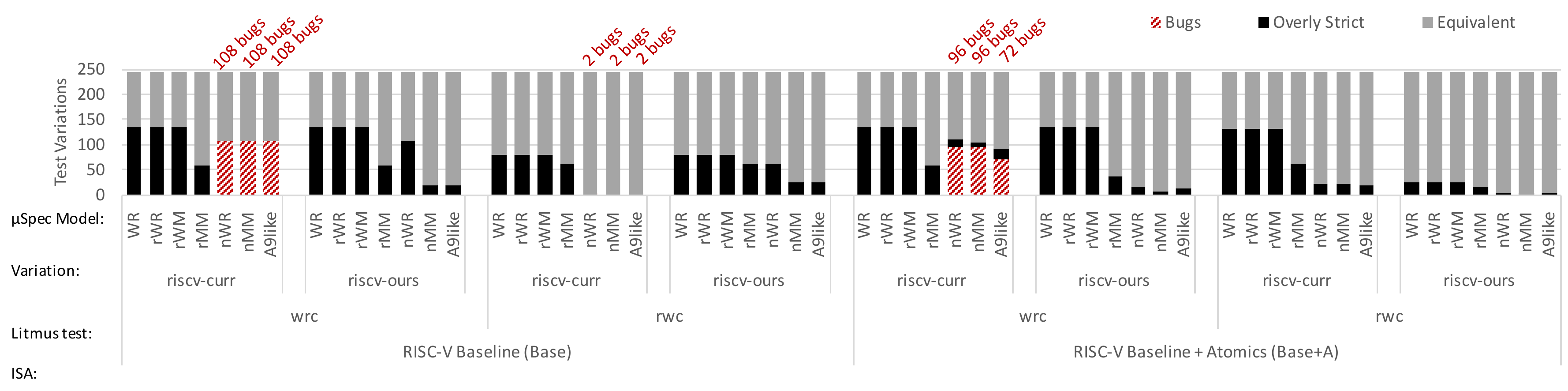}} \\
    {\label{fig:results81B}\includegraphics[width=.85\linewidth]{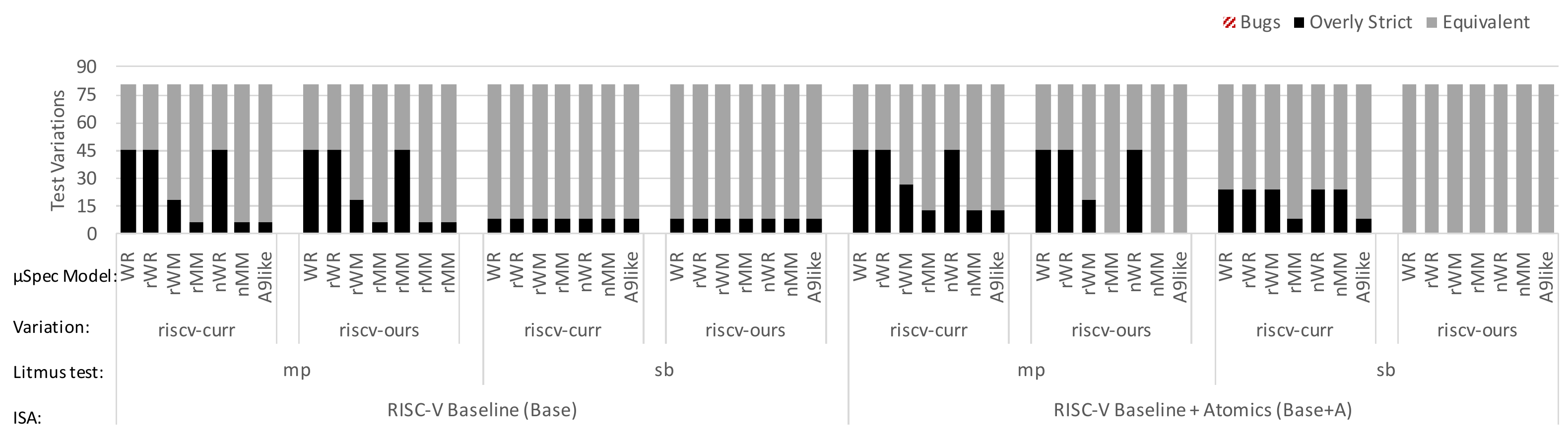}}  \\
    {\label{fig:results729A}\includegraphics[width=.41\linewidth]{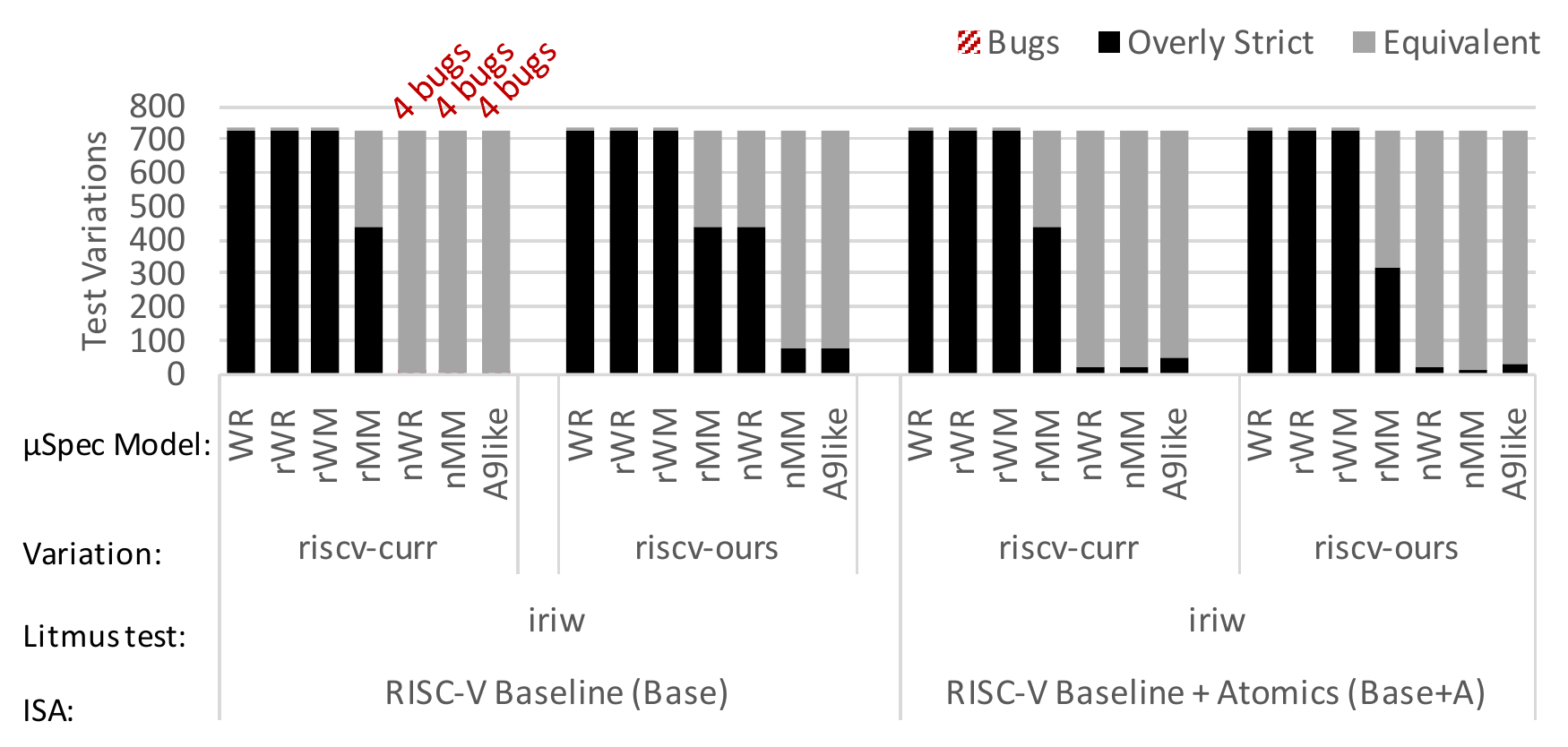}
     \label{fig:results729B}\includegraphics[width=.405\linewidth]{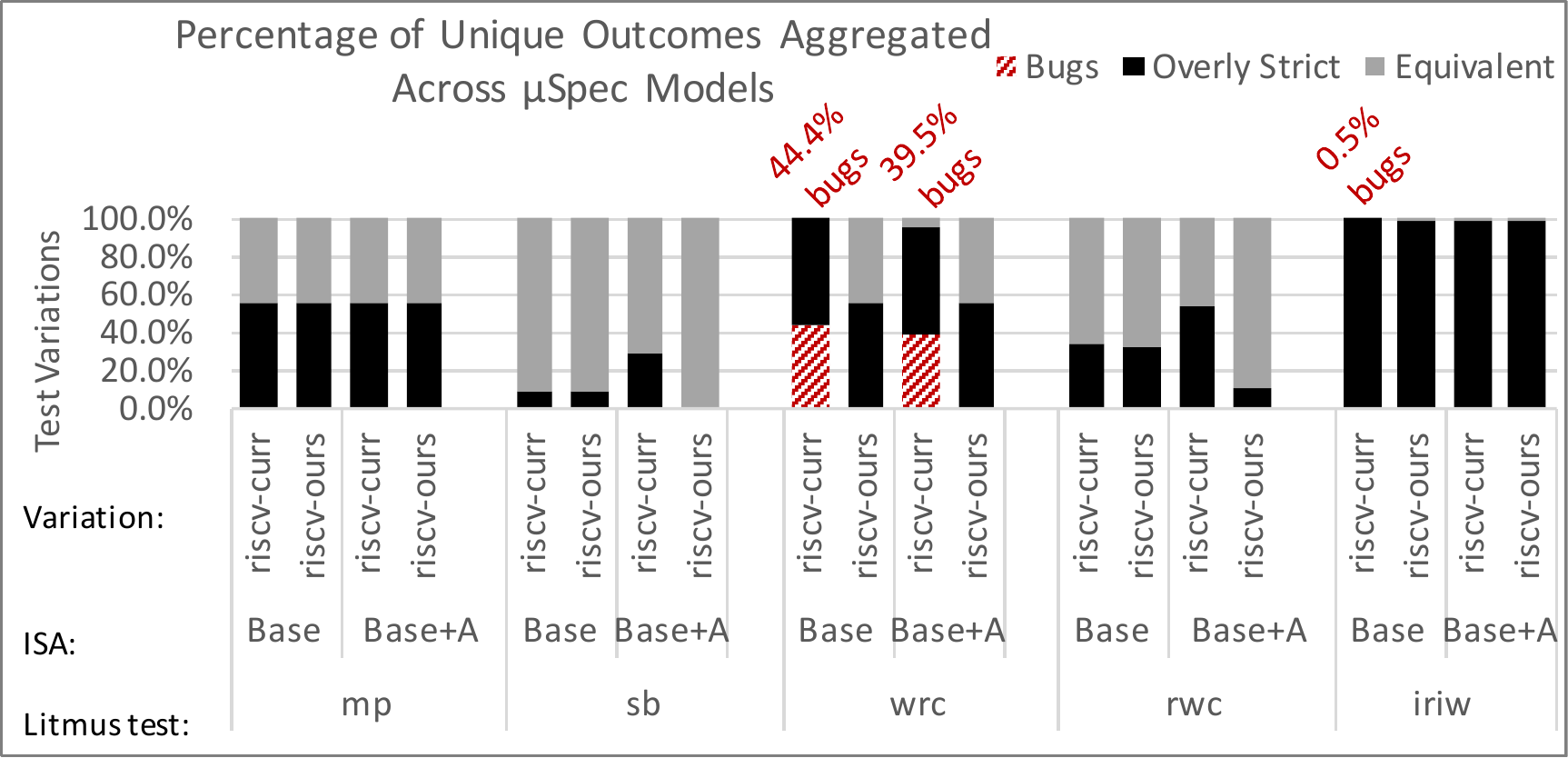}}
    \caption{Results of Step~\ref{item:equivalence} of the TriCheck methodology (Figure~\ref{fig:method}) for the Base and Base+A RISC-V MCMs. \textit{riscv-curr} and \textit{riscv-ours} refer to the \textit{current} RISC-V MCM~\cite{RISCV} and \textit{refined} version, respectively. Red striped bars are the number of executions that are \textit{forbidden} by the C11 HLL MCM, yet \textit{observable} for the tests in the input set. Black bars are the number of executions that are \textit{permitted} by the C11 HLL MCM, yet \textit{unobservable} for the tests in the input set.  Gray bars are tests that behave exactly as the C11 MCM indicates.}
    \label{fig:results}
\end{figure*}

As laid out in Section~\ref{sec:riscv}, Table~\ref{table:relaxations}, we evaluated a range of RISC-V microarchitectures, based off the Rocket Chip~\cite{rocketchip}, featuring a diverse set of RISC-V compliant memory order relaxations.
These were mapped with the appropriate RISC-V instructions depending on whether they were implementing the Base or Base+A ISA.
For each of the Base and Base+A MCMs, Figure~\ref{fig:results} shows results for \textit{riscv-curr} and \textit{riscv-ours} versions as inputs to our toolflow. 
The \textit{riscv-curr} version of the Base (resp.\@ Base+A) MCM corresponds to the initial set of inputs to our toolflow: \textit{current} Base (resp. Base+A) RISC-V MCM~\cite{RISCV}, ``Intuitive'' compiler mappings of Table~\ref{table:compilermappings} (resp. Table~\ref{table:compilermappingsAMO}), and Base (resp.\@ Base+A) RISC-V implementations of the Table~\ref{table:relaxations} $\mu$Spec models.
The \textit{riscv-ours} version of the Base (resp.\@ Base+A) MCM corresponds to the final results of the refinement process of Section~\ref{sec:casestudy}: \textit{refined} Base (resp.\@ Base+A) RISC-V MCM, ``Refined'' compiler mappings of Table~\ref{table:compilermappings} (resp. Table~\ref{table:compilermappingsAMO}), and \textit{refined} Base (resp. Base+A) RISC-V implementations of the Table~\ref{table:relaxations} $\mu$Spec models to accommodate RISC-V MCM changes.

The chart in the bottom right-hand corner of Figure~\ref{fig:results} additionally depicts results aggregated across litmus tests in each litmus test suite. Bug bars correspond to the percentage of tests that \textit{ever} produced an illegal outcome for a litmus test variation of a specified type when executed on \textit{any} of our microarchitectural implementations. The Overly Strict display the percentage of tests that \textit{ever} produced an Overly Strict outcome, but \textit{never} a Bug. Equivalent bars are the percentage of tests that always produced C11-specified outcomes.

\subsection{Litmus Test Suite Evaluation}
In Section~\ref{sec:motivating}, we discussed MCM features and alluded to issues that can result when these features are not carefully taken into account at design time. Through our Section~\ref{sec:casestudy} case study, we found that the Base and Base+A RISC-V MCMs are prone to pitfalls via these same MCM features. 
All of these errors enumerated below are eliminated in our refined riscv-ours $\mu$Spec model, ISA MCM, and compiler mappings, for both RISC-V Base and Base+A.

\textbf{Lack of Cumulative Lightweight Fences from Section~\ref{sec:base_lack_cum}}:
The $\mu$Spec models that are subject to errors as a result of the RISC-V MCM omitting cumulative lightweight fences are those with nMCA stores---nWR, nMM, and A9like.
A lack of cumulative lightweight fences in the Base riscv-curr versions of these nMCA models resulted in 108 illegal outcomes out of the 243 variants of the WRC litmus test.

\textbf{Lack of Cumulative Heavyweight Fences from Section~\ref{sec:base_lack_strong_cum}}:
Also applicable to the three nMCA $\mu$Spec models, is the omission of cumulative heavyweight fences from the RISC-V MCM. 
The result of this can be seen in the Base riscv-curr versions of the three nMCA microarchitectures for the RWC and IRIW litmus tests.
Each model exhibited 2 illegal outcomes out of the 243 variants of RWC.
Out of the 729 variations of IRIW, the nWR, nWW, and A9like models experienced 4 buggy executions.



\textbf{Reordering Loads to the Same Address from Section~\ref{sec:ld_ld_reorder_riscv}}: 
We observed read$\rightarrow$read reordering of reads of the same address on both the Base and Base+A RISC-V ISAs for the CoRR and CO-RSDWI litmus tests.
We do not include quantitative results for these tests in Figure~\ref{fig:results}, as they do not rely on any subtle interplay of instructions and are straightforwardly observable (yet forbidden by C11) when same-address loads are implemented with RISC-V relaxed loads.
For the Base and Base+A riscv-curr versions, CoRR variants produced 18 illegal results out of 81 for the $\mu$Spec models that relax read$\rightarrow$read ordering---rMM, nMM, and A9like.
CO-RSDWI variants produced 54 illegal results out of 243 for the same $\mu$Spec models.

\textbf{Lack of Cumulative Releases from Section~\ref{sec:lack_cum_releases}}:
The lack of cumulative releases in RISC-V again affects only nMCA implementations---rWR, rMM, and A9like---as displayed by the WRC executions for the Base+A riscv-curr versions of these $\mu$Spec models. Out of the 243 WRC variants, the rWR and rMM variants produce 96 illegal outcomes, and A9like exhibits 72. 

Note that the RWC and IRIW litmus tests are only forbidden at the C11 level when SC atomics are involved. Thus, the non-cumulative behavior of riscv-curr acquires and releases is not buggy for these tests unless SC atomics are used, resulting in fewer cases being flagged as bugs.


\textbf{Absence of Roach-Motel Movement for SC Atomics from Section~\ref{sec:roach_motel}}:
The effects of this on overly-constraining C11 programs can be seen by comparing all Base+A riscv-curr and riscv-ours variants and noting that the Overly Strict bars decrease in size from riscv-curr to riscv-ours (or stay the same in a couple of cases). When they stay the same, e.g. IRIW, this is because the microarchitectures themselves are not relaxed enough to exploit the difference between SC operations that allow roach-motel and those that don't.

%% file: 07-future.tex
\section{Applying TriCheck to Compiler Evaluation}
\label{sec:future}
As discussed in Section~\ref{sec:uarch_explanation}, we evaluated a microarchitecture very similar in ordering semantics to those allowed by the ARM Cortex-A9 (specifically, the A9like microarchitecture) when conducting our RISC-V case study. Two well-known compiler mappings from C11 to the Power and ARMv7 architectures are the \textit{leading-sync} mapping~\cite{mckenney} and the \textit{trailing-sync} mapping~\cite{c++topower}, both of which were supposedly proven correct~\cite{c++topower}. We initially elected to use the trailing-sync compiler mapping for our analysis. In doing so, TriCheck identified two counter-examples to this mapping on the A9like microarchitecture, thus invalidating the mappings. This led to our discovery of a loophole~\cite{manerkar:compilermappings} in the compilation proof which had allowed the incorrect mappings to pass through. We decided instead to use the leading-sync mapping for our analysis, as reflected in Table~\ref{table:compilermappings}.

Concurrent work identified a counter-example to the \textit{leading-sync} mapping as well~\cite{dryer:repairingc++}. This counter-example makes use of C11 SC fences. Since we did not evaluate the mixing of C11 fences and atomic instructions in this work, we did not observe this bug in our case study. The presence of counter-examples for both leading and trailing-sync mappings leaves C11 without a provably-correct mapping to Power and ARMv7 at present. Ongoing work has proposed a weakening of the C11 MCM specification to fix this problem~\cite{dryer:repairingc++}, but the situation remains in flux. This example demonstrates TriCheck's applicability beyond this paper's main focus on ISA design. 

%% file: 08-related.tex
\section{Related Work}
\label{sec:related}

In the past decade, researchers have formalized the specifications of a number of important real-world memory models.
Java, x86-TSO, Power, ARM, C11, and OpenCL have been formalized operationally~\cite{owens:better,petri:cooking,sarkar2011} and/or axiomatically~\cite{alglave:herd,batty:overhauling,mador-haim:axiomatic,commoncompiler}.
Most of these efforts, however, use some pre-existing document(s) as a starting point, and generally the refinement is performed according to the designers' original intent.
In contrast, this work treats the software requirements, microarchitectural guarantees, and ISA MCM speicifiactions as design parameters than can be explored and modified.

{\bf\noindent Verifying HLL Mappings onto Weak ISA MCMs: \hspace{0.1in}}
The two programming languages that have received the most attention in terms of memory model formalization are C11 and Java.
In a series of work, Batty et al.\@ developed a mathematically rigorous semantics for C11 concurrency, formalized using the Isabelle/HOL theorem prover via Lem~\cite{batty:overhauling,Batty:mathematizingc++,batty:lem,isabellehol}.
As part of this process, they produced a verified compilation scheme from C11/C++11 onto the x86, ARM, and Power MCMs~\cite{c++topower,POWERlrsc}.
Vafeiadis et al.\@ developed various methods for proving the correctness of operations performed within a C11 compiler~\cite{commoncompiler,Vafeiadis:programlogicc11}.
Petri et al.\@ developed an operational model of Java which specifically focused on its mapping onto x86 and Power~\cite{petri:cooking}.
Mappings from HLLs onto other architectures have also been considered with varying degrees of formality~\cite{cambridge}.

{\bf\noindent Verifying Microarchitectures against ISA MCMs: \hspace{0.1in}}
Prior work has also enabled flexible verification of hardware with respect to its ISA-level memory consistency model specification.
Lustig et al.\@ and Manerkar et al.\@ developed a set of tools for specifying memory ordering behavior at the microarchitecture level and then comparing it to the ISA specification~\cite{pipecheck,coatcheck,ccicheck}.
We use these Check tools in this work.
Another Check tool is capable of injecting ordering enforcement mechanisms to restore orderings required by the ISA but not implemented (or incorrectly implemented) in hardware~\cite{armor}.
Finally, extensive work has developed black-box testing methodologies using litmus tests~\cite{hangal:tsotool}.
We draw from these techniques and expand on them in TriCheck.


%% file: 09-conclusion.tex
\section{Conclusion}
\label{sec:conclusion}
MCM design choices are complicated and involve reasoning about the subtle interplay between many diverse features.
Modifications to any layer in the hardware-software stack may expose inefficiencies or inaccuracies within the specification. Full-stack verification with TriCheck is necessary to ensure that HLL, compiler, ISA, and implementation align well on MCM requirements. In our RISC-V MCM case study, we
found that one RISC-V-compliant
microachitecture allows 144 outcomes forbidden by C11
to be observed out of 1,701 litmus tests examined. We also
demonstrated, however, that the same issues were not present across all RISC-V-compliant hardware designs.
Using TriCheck, ISA designers can iteratively refine and evaluate ISA specifications in a  microarchitecture-aware manner based on the ISA's ability to serve as a target for compiled C11 programs.


%% file: 10-ack.tex
\section{Acknowledgements}
\label{sec:ack}
We thank the anonymous reviewers for their helpful feedback.
This work was supported in part by C-FAR (under the grant HR0011-13-3-0002), one of the six SRC STARnet Centers, sponsored by MARCO and DARPA, and in part by the National Science Foundation (under grants CCF- 1117147 and CCF-1253700).

%% file: main.bbl
\begin{thebibliography}{10}

\bibitem{adve:tutorial}
Sarita Adve and Kourosh Gharachorloo.
\newblock Shared memory consistency models: A tutorial.
\newblock {\em {IEEE} {C}omputer}, 29(12):66--76, 1996.

\bibitem{alglave:herd}
Jade Alglave, Luc Maranget, and Michael Tautschnig.
\newblock Herding cats: Modelling, simulation, testing, and data mining for
  weak memory.
\newblock {\em ACM Transactions on Programming Languages and Systems (TOPLAS)},
  36(2):7:1--7:74, July 2014.

\bibitem{armcortexa9}
{ARM}.
\newblock {ARM} {Cortex}-{A9} technical reference manual {ARMv7}-{A},
  2008-2012.
\newblock
  \url{http://infocenter.arm.com/help/topic/com.arm.doc.ddi0388i/DDI0388I_cortex_a9_r4p1_trm.pdf}.

\bibitem{ARMHazard}
ARM.
\newblock Cortex-{A9} {MPCore}, programmer advice notice, read-after-read
  hazards. {ARM} {Reference} 761319., 2011.
\newblock
  \url{http://infocenter.arm.com/help/topic/com.arm.doc.uan0004a/UAN0004A_a9_read_read.pdf}.

\bibitem{rocketchip}
Krste Asanovic, Rimas Avizienis, Jonathan Bachrach, Scott Beamer, David
  Biancolin, Christopher Celio, Henry Cook, Daniel Dabbelt, John Hauser, Adam
  Izraelevitz, Sagar Karandikar, Ben Keller, Donggyu Kim, John Koenig, Yunsup
  Lee, Eric Love, Martin Maas, Albert Magyar, Howard Mao, Miquel Moreto, Albert
  Ou, David~A. Patterson, Brian Richards, Colin Schmidt, Stephen Twigg, Huy Vo,
  and Andrew Waterman.
\newblock The {Rocket} {Chip} generator.
\newblock Technical Report UCB/EECS-2016-17, EECS Department, University of
  California, Berkeley, Apr 2016.

\bibitem{batty:overhauling}
Mark Batty, Alastair~F. Donaldson, and John Wickerson.
\newblock Overhauling {SC} atomics in {C11} and {OpenCL}.
\newblock In {\em 43rd Annual Symposium on Principles of Programming Languages
  (POPL)}, 2016.

\bibitem{problemconcurrency}
Mark Batty, Kayvan Memarian, Kyndylan Nienhuis, Jean Pichon-Pharabod, and Peter
  Sewell.
\newblock The problem of programming language concurrency semantics.
\newblock In {\em 24th European Symposium on Programming (ESOP), part of the
  European Joint Conferences on Theory and Practice of Software (ETAPS)}, 2015.

\bibitem{c++topower}
Mark Batty, Kayvan Memarian, Scott Owens, Susmit Sarkar, and Peter Sewell.
\newblock Clarifying and compiling {C}/{C++} concurrency: From {C++11} to
  {POWER}.
\newblock In {\em 39th Annual Symposium on Principles of Programming Languages
  (POPL)}, 2012.

\bibitem{Batty:mathematizingc++}
Mark Batty, Scott Owens, Susmit Sarkar, Peter Sewell, and Tjark Weber.
\newblock Mathematizing {C++} concurrency.
\newblock In {\em 38th Annual Symposium on Principles of Programming Languages
  (POPL)}, 2011.

\bibitem{invisifence}
Colin Blundell, Milo~M.K. Martin, and Thomas~F. Wenisch.
\newblock {InvisiFence}: Performance-transparent memory ordering in
  conventional multiprocessors.
\newblock In {\em 36th Annual International Symposium on Computer Architecture
  (ISCA)}, 2009.

\bibitem{Boehm05}
Hans-J. Boehm.
\newblock Threads cannot be implemented as a library.
\newblock In {\em Proceedings of the 2005 ACM SIGPLAN Conference on Programming
  Language Design and Implementation}, PLDI '05, pages 261--268, New York, NY,
  USA, 2005. ACM.

\bibitem{cppconcurrency}
Hans-J. Boehm and Sarita~V. Adve.
\newblock Foundations of the {C++} concurrency memory model.
\newblock In {\em 29th Conference on Programming Language Design and
  Implementation (PLDI)}, 2008.

\bibitem{bulksc}
Luis Ceze, James Tuck, Pablo Montesinos, and Josep Torrellas.
\newblock {BulkSC}: Bulk enforcement of sequential consistency.
\newblock In {\em 34th Annual International Symposium on Computer Architecture
  (ISCA)}, 2007.

\bibitem{collier}
William~W. Collier.
\newblock {\em Reasoning About Parallel Architectures}.
\newblock Prentice-Hall, Inc., Upper Saddle River, NJ, USA, 1992.

\bibitem{elver14}
M.~Elver and V.~Nagarajan.
\newblock {TSO}-{CC}: Consistency directed cache coherence for {TSO}.
\newblock In {\em 20th International Symposium on High Performance Computer
  Architecture (HPCA)}, 2014.

\bibitem{gharachorloo:thesis}
Kourosh Gharachorloo.
\newblock {\em Memory Consistency Models for Shared-memory Multiprocessors}.
\newblock PhD thesis, Stanford University, Stanford, CA, USA, 1996.

\bibitem{gharachorloo:release}
Kourosh Gharachorloo, Daniel Lenoski, James Laudon, Phillip Gibbons, Anoop
  Gupta, and John Hennessy.
\newblock Memory consistency and event ordering in scalable shared-memory
  multiprocessors.
\newblock {\em 17th International Symposium on Computer Architecture (ISCA)},
  1990.

\bibitem{speculativesc}
Chris Gniady and Babak Falsafi.
\newblock Speculative sequential consistency with little custom storage.
\newblock In {\em International Conference on Parallel Architectures and
  Compilation Techniques (PACT)}, 2002.

\bibitem{sc_ilp_rc}
Chris Gniady, Babak Falsafi, and T.N. Vijaykumar.
\newblock Is {SC} + {ILP} = {RC}?
\newblock {\em 41st International Symposium on Computer Architecture (ISCA)},
  1999.

\bibitem{atomicsc}
Dibakar Gope and Mikko~H. Lipasti.
\newblock Atomic {SC} for simple in-order processors.
\newblock In {\em 20th International Symposium on High Performance Computer
  Architecture {HPCA}}, 2014.

\bibitem{checkweb}
Martonosi~Research Group.
\newblock Check research tools and papers website, 2017.
\newblock http://check.cs.princeton.edu.

\bibitem{hangal:tsotool}
Sudheendra Hangal, Durgam Vahia, Chaiyasit Manovit, and Juin-Yeu~Joseph Lu.
\newblock {TSOtool}: A program for verifying memory systems using the memory
  consistency model.
\newblock In {\em 31st Annual International Symposium on Computer Architecture
  (ISCA)}, 2004.

\bibitem{hower14}
Derek~R. Hower, Blake~A. Hechtman, Bradford~M. Beckmann, Benedict~R. Gaster,
  Mark~D. Hill, Steven~K. Reinhardt, and David~A. Wood.
\newblock Heterogeneous-race-free memory models.
\newblock In {\em 19th International Conference on Architectural Support for
  Programming Languages and Operating Systems (ASPLOS)}, 2014.

\bibitem{cpp14}
ISO/IEC.
\newblock {Programming Languages -- C++}, 2014.

\bibitem{jackson:software}
Daniel Jackson.
\newblock {\em Software Abstractions: logic, language, and analysis}.
\newblock MIT Press, 2012.

\bibitem{keleher92}
Pete Keleher, Alan~L. Cox, and Willy Zwaenepoel.
\newblock Lazy release consistency for software distributed shared memory.
\newblock In {\em 19th Annual International Symposium on Computer
  Architecture}, 1992.

\bibitem{dryer:repairingc++}
Ori Lahav, Viktor Vafeiadis, Jeehoon Kang, Chung-Kil Hur, and Derek Dreyer.
\newblock Repairing sequential consistency in {C/C++11}.
\newblock {\em MPI-SWS}, Tech. rep. MPI-SWS-2016-011, 2016.

\bibitem{lamport:sc}
Leslie Lamport.
\newblock How to make a multiprocessor computer that correctly executes
  multiprocess programs.
\newblock {\em IEEE Transactions on Computing}, 28(9):690--691, 1979.

\bibitem{efficientsc}
Changhui Lin, Vijay Nagarajan, Rajiv Gupta, and Bharghava Rajaram.
\newblock Efficient sequential consistency via conflict ordering.
\newblock In {\em 17th International Conference on Architectural Support for
  Programming Languages and Operating Systems (ASPLOS)}, 2012.

\bibitem{pipecheck}
Daniel Lustig, Michael Pellauer, and Margaret Martonosi.
\newblock {PipeCheck}: Specifying and verifying microarchitectural enforcement
  of memory consistency models.
\newblock In {\em 47th International Symposium on Microarchitecture (MICRO)},
  2014.

\bibitem{coatcheck}
Daniel Lustig, Geet Sethi, Margaret Martonosi, and Abhishek Bhattacharjee.
\newblock {COATCheck: Verifying Memory Ordering at the Hardware-OS Interface}.
\newblock In {\em Proceedings of the 21st International Conference on
  Architectural Support for Programming Languages and Operating Systems}, 2016.

\bibitem{armor}
Daniel Lustig, Caroline Trippel, Michael Pellauer, and Margaret Martonosi.
\newblock {ArMOR}: Defending against memory consistency model mismatches in
  heterogeneous architectures.
\newblock In {\em 42nd International Symposium on Computer Architecture
  (ISCA)}, 2015.

\bibitem{lustig:automated}
Daniel Lustig, Andrew Wright, Alexandros Papakonstantinou, and Olivier Giroux.
\newblock Automated synthesis of comprehensive memory model litmus test suites.
\newblock {\em 22nd International Conference on Architectural Support for
  Programming Languages and Operating Systems (ASPLOS)}, 2017.

\bibitem{mador-haim:axiomatic}
Sela Mador-Haim, Luc Maranget, Susmit Sarkar, Kayvan Memarian, Jade Alglave,
  Scott Owens, Rajeev Alur, Milo M.~K. Martin, Peter Sewell, and Derek
  Williams.
\newblock An axiomatic memory model for {POWER} multiprocessors.
\newblock In {\em 24th International Conference on Computer Aided Verification
  (CAV)}, 2012.

\bibitem{ccicheck}
Yatin~A. Manerkar, Daniel Lustig, Michael Pellauer, and Margaret Martonosi.
\newblock {CCICheck}: Using $\mu$hb graphs to verify the coherence-consistency
  interface.
\newblock In {\em 48th International Symposium on Microarchitecture (MICRO)},
  2015.

\bibitem{manerkar:compilermappings}
Yatin~A. Manerkar, Caroline Trippel, Daniel Lustig, Michael Pellauer, and
  Margaret Martonosi.
\newblock Counterexamples and proof loophole for the {C/C++} to {POWER} and
  armv7 trailing-sync compiler mappings.
\newblock {\em CoRR}, abs/1611.01507, 2016.

\bibitem{Martin:valueprediction}
Milo M.~K. Martin, Daniel~J. Sorin, Harold~W. Cain, Mark~D. Hill, and Mikko~H.
  Lipasti.
\newblock Correctly implementing value prediction in microprocessors that
  support multithreading or multiprocessing.
\newblock In {\em 34th International Symposium on Microarchitecture (MICRO)},
  2001.

\bibitem{mckenney}
Paul~E. McKenney and Raul Silvera.
\newblock Example {POWER} implementation for {C/C++} memory model, 2011.
\newblock
  \url{http://www.rdrop.com/users/paulmck/scalability/paper/N2745r.2011.03.04a.html}.

\bibitem{batty:lem}
Dominic~P. Mulligan, Scott Owens, Kathryn~E. Gray, Tom Ridge, and Peter Sewell.
\newblock Lem: Reusable engineering of real-world semantics.
\newblock In {\em 19th International Conference on Functional Programming
  (ICFP)}, 2014.

\bibitem{isabellehol}
Tobias Nipkow, Markus Wenzel, and Lawrence~C. Paulson.
\newblock {\em Isabelle/{HOL}: A Proof Assistant for Higher-order Logic}.
\newblock Springer-Verlag, Berlin, Heidelberg, 2002.

\bibitem{owens:better}
Scott Owens, Susmit Sarkar, and Peter Sewell.
\newblock A better x86 memory model: {x86-TSO}.
\newblock In {\em 22nd International Conference on Theorem Proving in Higher
  Order Logics (TPHOLs)}, 2009.

\bibitem{petri:cooking}
Gustavo Petri, Jan Vitek, and Suresh Jagannathan.
\newblock Cooking the books: Formalizing {JMM} implementation recipes.
\newblock In {\em 29th European Conference on Object-Oriented Programming
  (ECOOP)}, 2015.

\bibitem{Ranganathan:speculativeretirement}
Parthasarathy Ranganathan, Vijay~S. Pai, and Sarita~V. Adve.
\newblock Using speculative retirement and larger instruction windows to narrow
  the performance gap between memory consistency models.
\newblock In {\em 9th Symposium on Parallel Algorithms and Architectures
  (SPAA)}, 1997.

\bibitem{linuxriscvbarrier}
{RISC-V Foundation}.
\newblock {RISC}-{V} port of {Linux} kernel, 2016.
\newblock
  \url{https://github.com/riscv/riscv-linux/blob/master/arch/riscv/include/asm/barrier.h}.

\bibitem{POWERlrsc}
Susmit Sarkar, Kayvan Memarian, Scott Owens, Mark Batty, Peter Sewell, Luc
  Maranget, Jade Alglave, and Derek Williams.
\newblock Synchronising {C}/{C++} and {POWER}.
\newblock In {\em 33rd Conference on Programming Language Design and
  Implementation (PLDI)}, 2012.

\bibitem{sarkar2011}
Susmit Sarkar, Peter Sewell, Jade Alglave, Luc Maranget, and Derek Williams.
\newblock Understanding power multiprocessors.
\newblock In {\em Proceedings of the 32Nd ACM SIGPLAN Conference on Programming
  Language Design and Implementation}, PLDI '11, pages 175--186, New York, NY,
  USA, 2011. ACM.

\bibitem{sewell:mappings}
Peter Sewell.
\newblock C/c++11 mappings to processors.
\newblock 2016.

\bibitem{cambridge}
Peter Sewell et~al.
\newblock C/{C++11} mappings to processors, 2016.
\newblock \url{https://www.cl.cam.ac.uk/~pes20/cpp/cpp0xmappings.html}.

\bibitem{endtoendsc}
Abhayendra Singh, Satish Narayanasamy, Daniel Marino, Todd Millstein, and
  Madanlal Musuvathi.
\newblock End-to-end sequential consistency.
\newblock In {\em 39th International Symposium on Computer Architecture
  (ISCA)}, 2012.

\bibitem{SPARCRMO}
{SPARC International}.
\newblock {\em The {SPARC} Architecture Manual (Version 9)}.
\newblock Prentice-Hall, Inc., Upper Saddle River, NJ, USA, 1994.

\bibitem{POWER4}
J.~M. Tendler, J.~S. Dodson, J.~S. Fields, H.~Le, and B.~Sinharoy.
\newblock {POWER4} system microarchitecture.
\newblock {\em IBM Journal of Research and Development}, 46(1):5--25, January
  2002.

\bibitem{linuxbarrier}
Linus Torvalds et~al.
\newblock Linux kernel, 2016.
\newblock
  \url{https://github.com/torvalds/linux/blob/master/arch/alpha/include/asm/barrier.h}.

\bibitem{commoncompiler}
Viktor Vafeiadis, Thibaut Balabonski, Soham Chakraborty, Robin Morisset, and
  Francesco Zappa~Nardelli.
\newblock Common compiler optimisations are invalid in the {C11} memory model
  and what we can do about it.
\newblock In {\em 42nd Symposium on Principles of Programming Languages
  (POPL)}, 2015.

\bibitem{Vafeiadis:programlogicc11}
Viktor Vafeiadis and Chinmay Narayan.
\newblock Relaxed separation logic: A program logic for {C11} concurrency.
\newblock In {\em 28th International Conference on Object Oriented Programming
  Systems Languages and Applications (OOPSLA)}, 2013.

\bibitem{RISCV}
Andrew Waterman, Yunsup Lee, David~A. Patterson, and Krste Asanovic.
\newblock The {RISC}-{V} instruction set manual, volume {I}: User-level {ISA},
  version 2.1.
\newblock Technical Report UCB/EECS-2016-118, EECS Department, University of
  California, Berkeley, May 2016.

\bibitem{Wenisch:storewaitfree}
Thomas~F. Wenisch, Anastasia Ailamaki, Babak Falsafi, and Andreas Moshovos.
\newblock Mechanisms for store-wait-free multiprocessors.
\newblock In {\em 34th International Symposium on Computer Architecture
  (ISCA)}, 2007.

\bibitem{wickerson:comparing}
John Wickerson, Mark Batty, Tyler Sorensen, and George~A Constantinides.
\newblock Automatically comparing memory consistency models.
\newblock {\em 44th Symposium on Principles of Programming Languages (POPL)},
  2017.

\end{thebibliography}
